# Real-time fMRI neurofeedback of the mediodorsal and anterior thalamus enhances correlation between thalamic BOLD activity and alpha EEG rhythm

Vadim Zotev[1#], Masaya Misaki[1], Raquel Phillips[1], Chung Ki Wong[1], Jerzy Bodurka[1,2#]

[1]Laureate Institute for Brain Research, Tulsa, OK, USA;
[2]College of Engineering, Stephenson School of Biomedical Engineering, University of Oklahoma, Norman, OK, USA

**Abstract:** Real-time fMRI neurofeedback (rtfMRI-nf) with simultaneous EEG allows volitional modulation of BOLD activity of target brain regions and investigation of related electrophysiological activity. We applied this approach to study correlations between thalamic BOLD activity and alpha EEG rhythm. Healthy volunteers in the experimental group (EG, *n*=15) learned to upregulate BOLD activity of the target region consisting of the mediodorsal (MD) and anterior (AN) thalamic nuclei using rtfMRI-nf during retrieval of happy autobiographical memories. Healthy subjects in the control group (CG, *n*=14) were provided with a sham feedback. The EG participants were able to significantly increase BOLD activities of the MD and AN. Functional connectivity between the MD and the inferior precuneus was significantly enhanced during the rtfMRI-nf task. Average individual changes in the occipital alpha EEG power significantly correlated with the average MD BOLD activity levels for the EG. Temporal correlations between the occipital alpha EEG power and BOLD activities of the MD and AN were significantly enhanced, during the rtfMRI-nf task, for the EG compared to the CG. Temporal correlations with the alpha power were also significantly enhanced for the posterior nodes of the default mode network, including the precuneus/posterior cingulate, and for the dorsal striatum. Our findings suggest that the temporal correlation between the MD BOLD activity and posterior alpha EEG power is modulated by the interaction between the MD and the inferior precuneus, reflected in their functional connectivity. Our results demonstrate the potential of the rtfMRI-nf with simultaneous EEG for non-invasive neuromodulation studies of human brain function.

**Keywords:**
thalamus, alpha rhythm, memory, neurofeedback, real-time fMRI, EEG-fMRI, mediodorsal nucleus, anterior nucleus, precuneus, dorsal striatum

## INTRODUCTION

Understanding mechanisms of generation and modulation of alpha EEG rhythms is a fundamental problem in neuroscience that awaits its comprehensive solution. Despite considerable progress, such mechanisms remain poorly understood due to their complexity [Başar, 2012]. On the neuronal level, rhythmic electrical activity in the alpha band (8-13 Hz) in a normal brain arises from complex interactions among thalamocortical and corticocortical neuronal circuits [Hughes & Crunelli, 2005]. On the macroscopic level, alpha EEG rhythms reflect complex interplay between globally coherent and locally dominated dynamic processes, leading, respectively, to functional integration and functional specialization [Nunez et al., 2001].

Simultaneous EEG-fMRI studies [Mulert & Lemieux, 2010] employing resting-state paradigms have provided important insights into properties of alpha EEG activity, e.g. [de Munck et al., 2007; DiFrancesco et al., 2008; Feige et al., 2005; Goldman et al., 2002; Gonçalves et al., 2006; Laufs et al., 2003; Liu et al., 2012; Moosmann et al., 2003; Omata et al., 2013; Wu et al., 2010]. These studies have revealed negative temporal correlations between fluctuations in posterior alpha EEG power, convolved with the hemodynamic response function (HRF), and blood-oxygenation-level-dependent (BOLD) activity in the occipital cortex, particularly in the primary (V1) and secondary (V2) visual areas. Negative alpha-BOLD correlations are also observed in the visual areas of the thalamus, including the lateral geniculate nucleus (LGN) and the ventrolateral pulvinar [Liu et al., 2012]. In contrast, positive temporal correlations between posterior alpha EEG power and BOLD activity have been detected in the dorsal parts of the thalamus, usually within and around the mediodorsal nucleus (MD), e.g. [de Munck et al., 2007; DiFrancesco et al., 2008; Goldman et al., 2002; Liu et al., 2012; Omata et al., 2013; Wu et al., 2010]. Such positive correlations between metabolic activity in the thalamus and alpha EEG power, also observed in EEG-PET, are often cited as evidence of the important role played by the thalamus in generation and modulation of alpha rhythms, e.g. [Schreckenberger et al., 2004]. However, important questions remain. Scalp EEG has low sensitivity to neuronal activities in deep subcortical areas. Therefore, the positive alpha-BOLD correlations, observed for the MD, must be supported by cortical regions, interacting with the MD and participating in generation or modulation of posterior alpha rhythm. Resting EEG-fMRI studies have not been able to identify such regions. Further complicating interpretation is the fact that the MD (unlike the LGN) has no significant neuronal connections to the occipital cortex, and has limited neuronal connectivity to the parietal cortex [Kasdon & Jacobson, 1978; Selemon & Goldman-Rakic, 1988]. Spontaneous BOLD fluctuations in the MD and in the occipital/parietal regions appear to be uncorrelated in healthy subjects at rest [Allen et al., 2011; Kim et al., 2013]. Overall, underlying mechanisms of the positive alpha-BOLD correlations for the thalamus remain unknown.

[#]Corresponding authors. E-mail: vzotev@laureateinstitute.org
(V. Zotev), jbodurka@laureateinstitute.org (J. Bodurka)

Resting-state EEG-fMRI experimental paradigms have two limitations. *First*, they rely on spontaneous fluctuations in EEG and BOLD activities. This reduces the experiments' sensitivity to specific EEG-fMRI effects of interest. *Second*, a participant's actual mental state at any moment during rest cannot be controlled or measured accurately. This leads to large between-subject and within-subject variabilities in resting-state results. These limitations are evident in the above-mentioned resting EEG-fMRI studies of alpha rhythm, which showed both insufficient sensitivity, e.g. [Feige et al., 2005; Laufs et al., 2003; Moosmann et al., 2003], and large between-subject variability, e.g. [de Munck et al., 2007; Gonçalves et al., 2006].

In the present study, we investigate correlations between BOLD activity of the thalamus and posterior alpha EEG rhythm using the new multimodal approach we introduced recently – real-time fMRI neurofeedback with simultaneous EEG [Zotev et al., 2016]. Real-time fMRI neurofeedback (rtfMRI-nf), e.g. [Birbaumer et al., 2013; Thibault et al., 2016; Weiskopf, 2012] enables non-invasive volitional modulation of BOLD activities of small precisely defined regions deep inside the brain. The rtfMRI-nf modulation leads to an enhancement in functional connectivity between the target region and its task-specific network, e.g. [Zotev et al., 2011, 2013]. Electrophysiological activities of cortical regions belonging to the network can be probed using simultaneous (passive) scalp EEG recordings. Therefore, combination of the rtfMRI-nf and simultaneous EEG makes it possible to examine correlations between BOLD activity of the target region and related EEG activity [Zotev et al., 2016].

The rtfMRI-nf with simultaneous EEG approach has the potential to overcome the limitations inherent to resting-state EEG-fMRI. With rtfMRI-nf, BOLD activities of the target region and associated network can be significantly enhanced and modulated in a controlled manner. This increases signal-to-noise ratio and provides greater sensitivity for EEG-fMRI effects of interest. Because an rtfMRI-nf signal is a real-time measure of task performance, consistency of both individual and group results can be improved with sufficient training. Therefore, the rtfMRI-nf with simultaneous EEG offers new opportunities for studying relationships between electrophysiological and hemodynamic processes in the human brain. This approach can also help to identify promising target measures for EEG neurofeedback (EEG-nf), e.g. [Gruzelier, 2014], brain-computer interfaces, e.g. [Cavazza et al., 2014], and simultaneous multimodal rtfMRI-EEG-nf [Mano et al., 2017; Perronnet et al., 2017; Zotev et al., 2014].

For the present study, we selected the target region of interest (ROI) for rtfMRI-nf as the combination of the MD and the anterior nucleus (AN). The reason is that BOLD activity fluctuations in both the MD and the AN exhibit the strongest positive temporal correlations with posterior alpha EEG power at rest, according to the recent EEG-fMRI study [Liu et al., 2012]. Both the MD and AN are parts of the limbic thalamus [Taber et al., 2004; Wolff et al., 2015]. The MD has major reciprocal connections with the prefrontal cortex and the anterior cingulate cortex. It also receives inputs from the amygdala, basal forebrain, entorhinal cortex, temporal polar cortex, substantia nigra, and cerebellum [Mitchell, 2015; Taber et al., 2004]. Importantly, the MD subdivisions function as parts of the basal ganglia-thalamocortical circuits [Alexander et al., 1991]. The AN has major reciprocal connections with the anterior and posterior cingulate cortices, and the hippocampal complex. It also receives input from the mamillary complex [Child & Benarroch, 2013; Taber et al., 2004]. Both the MD and AN play important roles in memory, including episodic memory [Child & Benarroch, 2013; Taber et al., 2004; Wolff et al., 2015]. The MD also contributes to a variety of other brain functions, including emotion, motivation, executive function, learning, and decision-making [Mitchell, 2015; Taber et al., 2004].

Because the MD and AN thalamic nuclei are prominently involved in the episodic memory function, they are activated during recall of autobiographical memories, as demonstrated by meta-analyses of fMRI memory studies [Spreng et al., 2009; Svoboda et al., 2006]. We previously used an experimental paradigm involving retrieval of happy autobiographical memories together with rtfMRI-nf to achieve volitional modulation of the amygdala and associated emotion regulation network, including the MD [Zotev et al., 2011, 2016]. Importantly, episodic memory retrieval strongly activates the posterior nodes of the fMRI default mode network (DMN), including the precuneus / posterior cingulate and the left and right angular gyri [Sestieri et al., 2011; Spreng et al., 2009]. EEG source analysis studies have shown that the posterior DMN regions exhibit increased alpha EEG power during self-referential thoughts [Knyazev et al., 2011] and retrieval of autobiographical memories [Knyazev et al., 2015].

We conducted the rtfMRI-nf with simultaneous EEG experiment to test two main hypotheses. *First*, we hypothesized that healthy participants would be able to significantly increase BOLD activities of the MD and AN thalamic nuclei using the rtfMRI-nf during retrieval of happy autobiographical memories. *Second*, we hypothesized that performance of the rtfMRI-nf task would be accompanied by significant enhancements in temporal correlations between the MD and AN BOLD activities and posterior alpha EEG power. We also expected to identify cortical regions supporting these correlation effects.

## MATERIALS AND METHODS

### Participants

The study was conducted at the Laureate Institute for Brain Research, and was approved by the Western Institutional Review Board (IRB). All the participants provided written informed consent as approved by the IRB. All study procedures were performed in accordance with the principles expressed in the Declaration of Helsinki.

Thirty-four healthy volunteers participated in the study. The participants underwent a psychological evaluation by a licensed clinician and were both medically and psychiatrically healthy. Five subjects exhibited excessive head motions during scanning, and their data were not included in the analyses. The participants were randomly assigned to either the experimental group (EG, $n$=15, 9 females) or the control group (CG, $n$=14, 8 females). The EG participants followed an experimental procedure with rtfMRI-nf based on BOLD activity of the thalamus, as described in detail below. The CG participants followed the same procedure, but were provided, without their knowledge, with a sham feedback unrelated to any brain activity. The subjects' average age was 29 (*SD*=6) years for the EG and 29 (*SD*=10) years for the CG.

### Experimental Paradigm

The experimental paradigm involved targeted modulation of BOLD activity of the AN and MD thalamic nuclei using rtfMRI-nf during retrieval of happy autobiographical memories and recording of concurrent EEG activity (Figure 1).

The target ROI for the rtfMRI-nf for the EG consisted of the AN and MD nuclei, depicted in Fig. 1A. They were defined anatomically as the AN and MD regions specified in the stereotaxic atlas of the human brain by Talairach and Tournoux [Talairach & Tournoux, 1988]. The target ROI was transformed to each subject's individual fMRI image space and used as a mask to compute the average ongoing BOLD activity of the target region. The rtfMRI-nf signal provided to the EG participant was based on this activity. It was presented to the subject inside the MRI scanner in the form of a variable-height thermometer-style red bar on the screen, as described below. By controlling the height of the rtfMRI-nf bar, the EG participant was able to directly modulate BOLD activity of the target region in real time.

The sham feedback signal for the CG was computer-generated as illustrated in Fig. 1B. It was defined, for each 40-s long condition block, as a linear combination of seven Legendre polynomials with randomly selected coefficients, projected from the [−1...+1] interval onto the [0...40] s time interval. The random number generator was initialized to a random seed value at the beginning of each experiment. Such sham feedback signal definition yielded a smooth waveform (Fig. 1B), which was used to set the height of the red bar in real time. It appeared to provide meaningful real-time information to the CG participant. However, the waveform's shape (temporal profile) was random, and it also varied randomly across the condition blocks and across the subjects. We did not select the sham neurofeedback approach that employs an rtfMRI signal from a control brain region, e.g. [Zotev et al., 2011], because the thalamus has extensive neuronal connections to various cortical areas and subcortical structures.

### Experimental Protocol

Prior to the rtfMRI-nf session, each participant was given detailed instructions that included an overview of the experiment and an explanation of each experimental task. The participant was asked to think of and write down three happy autobiographical memories, keeping them private. It was suggested that he/she use those three memories at the beginning of the experiment to evaluate their effects, and then explore various other happy autobiographical memories as the training progressed to enhance emotional memory experience and improve rtfMRI-nf performance. The participants were encouraged to keep their eyes open and pay attention to the display screen throughout each experimental run.

The experimental protocol is illustrated in Fig. 2. It included three conditions: Happy Memories, Attend, and Count. The real-time GUI display screens for these conditions are shown in Fig. 2A. Each condition was specified by visual cues that included a color symbol at the center of the screen and a text line at the top. As mentioned above, the rtfMRI-nf signal was represented by the variable-height red bar (Fig. 2A). We chose to have it displayed and updated during all three experimental conditions to reduce variations in visual stimulation and visual attention levels across the conditions. The bar height was updated every 2 s, and was also indicated by the numeric value shown above the bar. The blue bar was fixed and denoted the zero level.

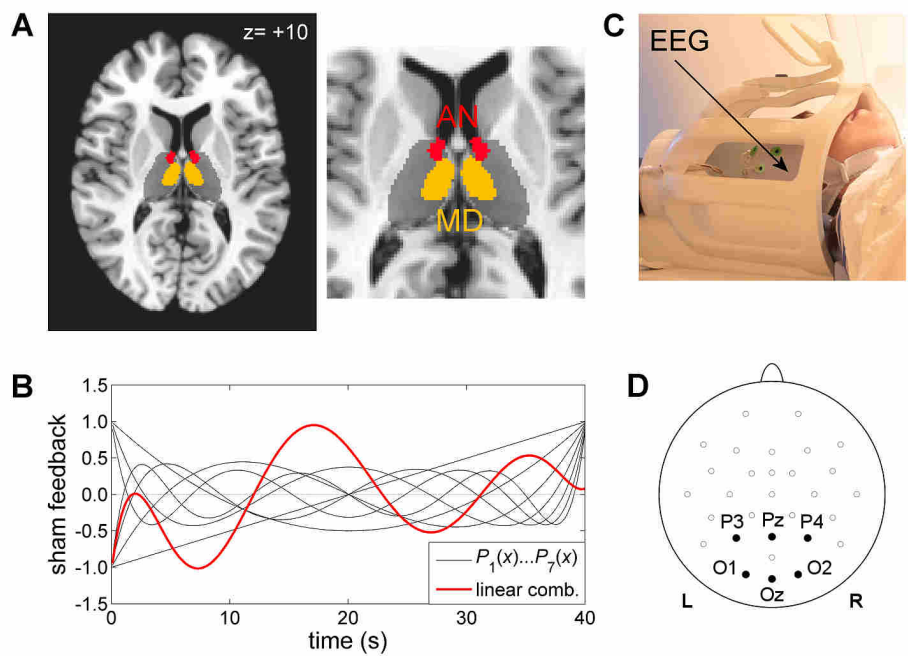

**Figure 1. Experimental paradigm for real-time fMRI neurofeedback modulation of the thalamus with simultaneous EEG recordings. A)** Target region of interest (ROI) used to provide the rtfMRI-nf signal for the experimental group (EG). The target ROI consists of two thalamic nuclei: the anterior nucleus (AN, red) and the mediodorsal nucleus (MD, orange). The nuclei are defined anatomically according to the stereotaxic atlas of the human brain by Talairach and Tournoux. They are projected in the figure onto the standard anatomical template TT_N27 in the Talairach space. The cross-section of the whole thalamus is shown in a darker grey color. **B)** Generation of the sham feedback signal for the control group (CG). The sham signal waveform is computed for a 40-s long condition block as a random linear combination of seven Legendre polynomials. **C)** A 32-channel MR-compatible EEG system was used to perform EEG recordings simultaneously with fMRI data acquisition. **D)** Occipital (O1, O2, Oz) and parietal (P3, P4, Pz) EEG channels commonly used to study posterior alpha EEG activity.

For the Happy Memories condition blocks, the participants were instructed to evoke and contemplate happy autobiographical memories while simultaneously trying to raise the red rtfMRI-nf bar on the screen as high as possible (Fig. 2A, left). For the Attend condition blocks, the subjects were asked to relax while paying attention to the fluctuating red bar without attempting to control it (Fig. 2A, middle). For the Count condition blocks, the participants were instructed to mentally count back from 300 by subtracting a given integer while watching the red bar without trying to control it (Fig. 2A, right). Because during the Happy Memories condition the subjects were asked to volitionally control the rtfMRI-nf signal, we also refer to it as the rtfMRI-nf task. The Attend condition served as a baseline attention task, and the Count condition was a cognitive control task.

The rtfMRI-nf experiment included seven fMRI runs (Fig. 2B) each lasting 8 min 46 s. During the initial and final Rest runs, the participants were asked to relax and rest while looking at the fixation cross. The five task runs – Run 1, Run 2, Run 3, Run 4, and the Transfer run – consisted of alternating 40-s long blocks of Happy Memories, Attend, and Count conditions (Fig. 2B). Each Happy Memories or Count condition block was preceded by an Attend condition block. During the four rtfMRI-nf runs (Runs 1-4), the participants performed the three experimental tasks as indicated by the GUI display with the rtfMRI-nf bar (Fig. 2A). During the Transfer run, the participants performed the same tasks, except that no bars were shown on the screen for any of the conditions. The Transfer run was included in order to evaluate whether the participants' learned ability to control BOLD activity of the target ROI generalized beyond the actual rtfMRI-nf training. The Happy Memories conditions did not differ across Runs 1-4. The Count conditions involved counting back from 300 by subtracting 3, 4, 6, 7, and 9 for Run 1, Run 2, Run 3, Run 4, and the Transfer run, respectively. The display range for the rtfMRI-nf signal was −2% to +2% for all conditions.

### Data Acquisition

All experiments were conducted on the General Electric Discovery MR750 3T MRI scanner with a standard 8-channel receive-only head coil (Fig. 1C). A single-shot gradient echo EPI sequence with FOV/slice=240/2.9 mm, *TR/TE*=2000/30 ms, flip angle=90°, 34 axial slices per volume, slice gap=0.5 mm, SENSE acceleration *R*=2 in the phase encoding (anterior-posterior) direction, acquisition matrix 96×96, sampling bandwidth=250 kHz, was employed for fMRI. Each fMRI run lasted 8 min 46 s and included 263 EPI volumes (the first three EPI volumes were excluded from data analysis to account for fMRI signal reaching a steady state). Physiological pulse oximetry and respiration waveforms were recorded simultaneously with fMRI. The EPI images were reconstructed into a 128×128 matrix, yielding 1.875×1.875×2.9 mm$^3$ fMRI voxels. A T1-weighted 3D MPRAGE se-quence with FOV/slice=240/1.2 mm, *TR*/*TE*=5.0/1.9 ms, *TD*/*TI*=1400/725 ms, flip angle=10°, 128 axial slices per slab, SENSE *R*=2, acquisition matrix 256×256, sampling bandwidth=31.2 kHz, scan time=4 min 58 s, was used for anatomical imaging. It provided structural brain images with 0.94×0.94×1.2 mm$^3$ voxel size.

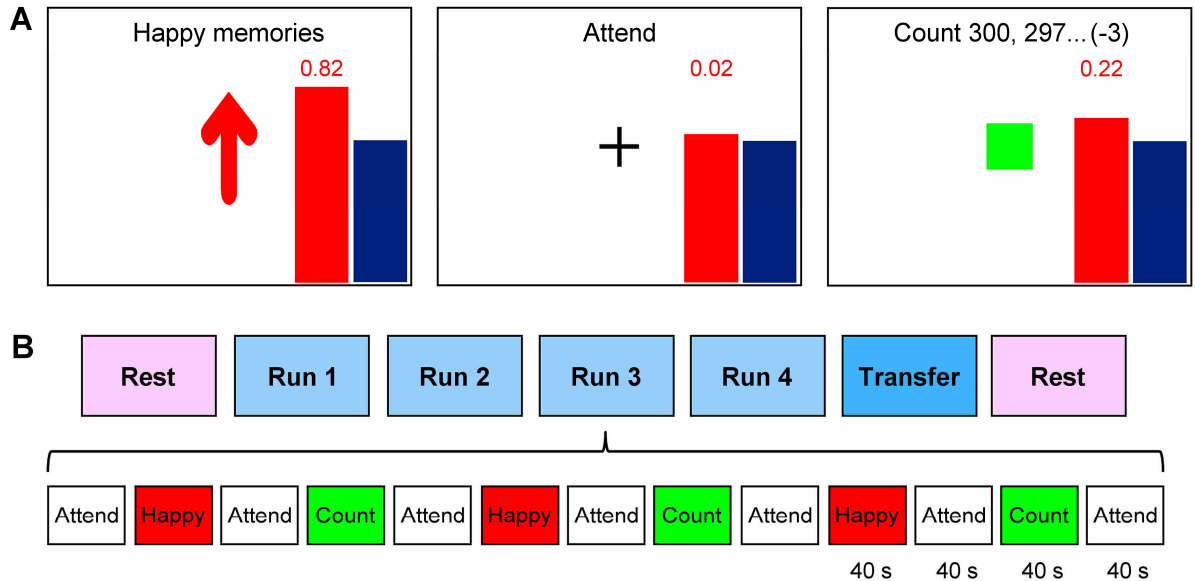

**Figure 2. Protocol for the rtfMRI-nf experiment. A)** Real-time GUI display screens for three experimental conditions: Happy Memories, Attend, and Count. The variable-height rtfMRI-nf bar (red) is shown during each condition, and its height is updated every 2 s. The fixed blue bar denotes the zero level. **B)** Protocol for the rtfMRI-nf experiment includes seven runs, each lasting 8 min 46 s: Rest (RE), Run 1 (R1), Run 2 (R2), Run 3 (R3), Run 4 (R4), Transfer (TR), and Rest (RE). The experimental runs (except the Rest) consist of 40-s long blocks of Happy Memories (H), Attend (A), and Count (C) conditions. No bars are shown during the Transfer run.

EEG recordings were performed simultaneously with fMRI (Fig. 1C) using a 32-channel MR-compatible EEG system from Brain Products, GmbH. The EEG system clock was synchronized with the MRI scanner 10 MHz clock using the Brain Products' SyncBox. EEG data were acquired with 0.2 ms temporal and 0.1 µV measurement resolution (16-bit 5 kS/s sampling) in 0.016...250 Hz frequency band with respect to FCz reference. All technical details of the EEG-fMRI system configuration and data acquisition were reported previously [Zotev et al., 2012]. Similar to our recent study [Zotev et al., 2016], the EEG recordings in the present work were passive, i.e. no EEG information was used in real time as part of the experimental procedure.

### Real-Time Data Processing

The rtfMRI-nf was implemented using the custom real-time fMRI system utilizing real-time functionality of AFNI [Cox, 1996; Cox & Hyde, 1997] as described previously [Zotev et al., 2011]. A high-resolution MPRAGE structural brain image and a short EPI dataset (5 volumes) were acquired at the beginning of each rtfMRI-nf experiment. The last volume of the EPI dataset was used as a reference EPI volume defining the subject's individual EPI space. The target ROI, defined in the Talairach space (Fig. 1A), was transformed to the individual EPI space using the MPRAGE image data. The resulting ROI in the EPI space contained, on the average, 260 voxels. During the subsequent fMRI runs (Fig. 2B), the AFNI real-time plugin was used to perform volume registration of each acquired EPI volume to the reference EPI volume (motion correction) and export the mean value of fMRI signal for the target ROI in real time. The custom developed GUI software was used to further process the exported fMRI signal values and display the ongoing rtfMRI-nf information (Fig. 2A).

The rtfMRI-nf signal was defined as the fMRI percent signal change with respect to a selected baseline. For each Happy Memories or Count condition, an average of fMRI signal values for the preceding 40-s long Attend condition block (Fig. 2B) was used as the baseline. For each Attend condition, the baseline was calculated as an average of fMRI signal values in the same Attend condition block acquired prior to the current fMRI signal value. Specifically, for the 1st volume in the Attend block, the percent signal change was set to zero; for the 2nd volume, the fMRI signal value from the 1st volume was used as the baseline; for the 3rd volume, the baseline was computed as an average of the fMRI signal values from the 1st and 2nd volumes; and so on. Thus, the rtfMRI-nf signal for the Attend conditions represented percent signal changes for fluctuating fMRI signal values with respect to the incrementally updated Attend baseline. To reduce effects of fMRI noise and physiological artifacts, a moving average of the current and two preceding rtfMRI percent-signal-change values [Zotev et al., 2011] from the same condition block was computed. (The recorded cardiac and respiratory waveforms were not used in the real-time processing). This moving average was used to set the height of the red rtfMRI-nf bar on the screen (Fig. 2A) for the EG. The height of the red bar and the numeric value above it were updated every $TR$=2 s. To provide sham feedback for the CG, the actual rtfMRI-nf values for the Happy Memories and Count conditions were substituted with the computer-generated sham feedback signal values (Fig. 1B). An example of the rtfMRI-nf signal time course for one EG participant is shown in *Supplementary Material* (Fig. S1).

### fMRI Data Analysis

Offline analysis of the fMRI data was performed in AFNI as described in detail in *Supplementary Material* (S1.1). It involved fMRI pre-processing with cardiorespiratory artifact correction [Glover et al., 2000], slice timing correction, and volume registration. A standard fMRI activation analysis for each task run involved solution of a general linear model (GLM) with the Happy Memories and Count block-stimulus regressors. Average GLM-based fMRI percent signal changes were then computed for the AN and MD ROIs (Fig. 1A), and for the LGN ROI (S1.1). A standard fMRI functional connectivity analysis with the MD ROI as the seed region was performed within GLM framework specifically for the Happy Memories conditions in each task run. The GLM also included time courses of fMRI motion parameters and average fMRI time courses of bilateral ROIs within white matter (WM) and ventricle cerebrospinal fluid (CSF) as nuisance co-variates [Jo et al., 2010]. An fMRI functional connectivity analysis with the AN ROI as the seed region was performed in a similar way (S1.1).

In addition to the standard fMRI activation and functional connectivity analyses, we conducted similar analyses (S1.1) that included the fMRI time course of the ROI encompassing the primary (V1, BA 17) and secondary (V2, BA 18) visual areas as a nuisance covariate. We refer to this ROI as V1/V2. BOLD activities of these areas exhibit strong negative correlations with posterior alpha EEG power, e.g. [Feige et al., 2005]. The inclusion of this additional covariate yielded fMRI results that were less dependent on variations in visual attention.

### EEG Data Processing

Offline processing of the EEG data, acquired during fMRI, was performed in BrainVision Analyzer 2.1 software as described in detail in *Supplementary Material* (S1.2). Removal of EEG artifacts was based on the average artifact subtraction and independent component analysis [Bell & Sejnowski, 1995; McMenamin et al., 2010]. Channel Cz was selected as a new reference. Following the artifact removal, the EEG data were downsampled to 8 ms temporal resolution.

The alpha EEG band was defined individually for each participant as [IAF−2...IAF+2] Hz, where IAF is the individual alpha peak frequency. The IAF was determined by inspection of average EEG spectra for the occipital and parietal EEG channels (Fig. 1D) across the Attend condition blocks in the four rtfMRI-nf runs (Fig. 2B).

### Alpha Envelope Correlation Analysis

To study amplitude modulation of alpha EEG activity across the EEG array, we conducted analyses of temporal correlations between alpha amplitude envelopes for all channel pairs. After the artifact removal, EEG signals from all channels (8-ms temporal resolution) were band-pass filtered (48 dB/octave) in the individual alpha band, defined above, to extract their alpha activities. The Hilbert transform was applied to obtain the analytic signal, and its magnitude defined the alpha envelope time course for each channel. Temporal correlation between the alpha envelopes for each pair of channels was analyzed for each experimental condition in each of the five task runs (Fig. 2B). Each analysis included a segmentation with 4.096 s intervals, a complex FFT with 0.244 Hz spectral resolution, and the Coherence transform implemented in Analyzer 2.1. The transform was used to compute a magnitude-squared correlation coefficient for two channels' time courses at a given frequency as a squared magnitude of their cross-covariance value normalized by their auto-covariance values at the same frequency. The results were averaged for the low frequency range [0.244...1.46] Hz, and a square root of the average was calculated. We refer to the resulting quantity as the alpha envelope correlation (AEC). The AEC was computed for each pair of EEG channels. Differences in the average AEC values between experimental conditions (Fig. 2B) were used to evaluate task-specific variations in correlated amplitude modulation of alpha rhythm across the EEG array.

### Occipital Alpha Power Analysis

To facilitate comparison of our EEG-fMRI results to those of the previous studies that investigated BOLD correlates of occipital alpha EEG power, e.g. [DiFrancesco et al., 2008; Feige et al., 2005; Goldman et al., 2002; Liu et al., 2012], we focused on alpha power for the occipital channels O1, O2, Oz (Fig. 1D). (All the analyses were later repeated for the parietal channels P3, P4, Pz). EEG signal power was computed using a continuous wavelet transform with Morlet wavelets for [0.25...15] Hz frequency range with 0.25 Hz frequency resolution and 8 ms temporal resolution. Normalized occipital alpha EEG power [Gasser et al., 1982] was defined for each time point as $\alpha O = \ln(P(O1)) + \ln(P(O2)) + \ln(P(Oz))$, where $P$ is the EEG signal power in the individual alpha band defined above. The normalized occipital alpha power values were averaged for 200-ms long time bins, linearly detrended, and converted to z-scores across each run. To account for possible modulating effects of eye movements [de Munck et al., 2007], including eye blinking and occasional eye closing, on the alpha EEG power, we orthogonalized the z-score time course for each run with respect to the corresponding EEG eye-blinking covariate (see *Supplementary Material* S1.3). We used the resulting z-score time courses, denoted as $z(\alpha O)$, to define EEG-based regressors for EEG-fMRI analyses. Average differences in the z-scores between experimental conditions were used to evaluate task-specific variations in the normalized occipital alpha EEG power.

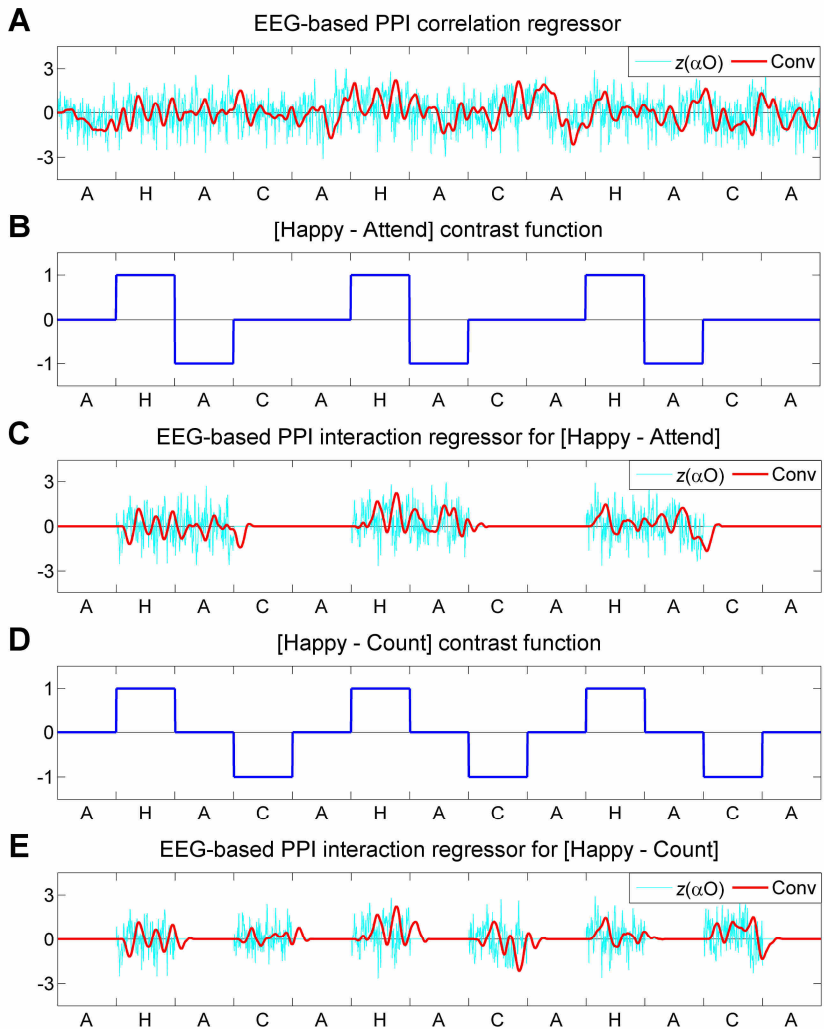

**Figure 3. Definition of EEG-based regressors for the psychophysiological interaction (PPI) analyses of EEG-fMRI data.** A general linear model (GLM) for a PPI analysis includes one PPI correlation regressor, one PPI interaction regressor, block-design stimulus regressors, and nuisance covariates. The PPI regressors are defined here using the time course of the normalized occipital alpha EEG power (αO) converted to z-scores across each run. **A)** Convolution of the EEG power z-scores (cyan, 200 ms sampling) with the hemodynamic response function (HRF) yields the EEG-based PPI correlation regressor (red). **B)** Definition of the [Happy−Attend] contrast function. It is equal to +1 for the Happy Memories (H) condition blocks, −1 for the following Attend (A) condition blocks, and 0 for all other points. The condition blocks are depicted schematically in Fig. 2B. **C)** Convolution of the EEG power z-scores multiplied by the [Happy−Attend] contrast function (cyan) with the HRF yields the EEG-based PPI interaction regressor for the Happy vs Attend condition contrast (red). **D)** Definition of the [Happy−Count] contrast function. It is equal to +1 for the Happy Memories (H) condition blocks, −1 for the Count (C) condition blocks, and 0 for all other points. **E)** Convolution of the EEG power z-scores multiplied by the [Happy−Count] contrast function (cyan) with the HRF yields the EEG-based PPI interaction regressor for the Happy vs Count condition contrast (red).

### EEG-fMRI Temporal Correlation Analysis

To study task-specific temporal correlations between alpha EEG power and BOLD activity, we performed psychophysiological interaction (PPI) analyses [Friston et al., 1997; Gitelman et al., 2003] adapted for EEG-fMRI [Zotev et al., 2014, 2016]. In these analyses, we tested the hypothesis that temporal correlations between the occipital alpha EEG power and BOLD activities of the AN and MD would be stronger during the Happy Memories conditions with rtfMRI-nf than during the control conditions (Attend, Count) for the EG, but not for the CG.

EEG-based regressors for the PPI analyses were defined as illustrated in Figure 3. The time course of z-scores of the normalized occipital alpha EEG power, $z(\alpha O)$, was computed for each run as described above. The HRF ('Cox special', maximum at 6 s) was calculated with 200 ms sampling using the waver AFNI program. Convolution of the $z(\alpha O)$ time course with the HRF produced the EEG-based regressor, employed as a PPI correlation regressor (Fig. 3A). In order to define a PPI interaction regressor, the $z(\alpha O)$ time course was first multiplied by a selected contrast function, and then convolved with the HRF. An EEG-based PPI interaction regressor for the Happy vs Attend condition contrast was defined as illustrated in Figs. 3B,C. An EEG-based PPI interaction regressor for the Happy vs Count condition contrast was defined as illustrated in Figs. 3D,E. These EEG-based PPI regressors were sub-sampled to middle time points of fMRI volumes, linearly detrended, and used as stimulus regressors in the PPI analyses of the EEG-fMRI data within GLM framework.

In addition to the EEG-based PPI regressors, we defined fMRI-based PPI regressors using the average fMRI time course of the visual cortex V1/V2, as described in *Supplementary Material* (S1.4). We refer to them as fMRI-visual PPI correlation covariate and fMRI-visual PPI interaction covariate (for the Happy vs Attend or Happy vs Count condition contrast). These waveforms were used as nuisance covariates in the PPI analyses of the EEG-fMRI data.

Each PPI analysis involved a solution of a GLM model with two EEG-based PPI regressors by means of the 3dDeconvolve AFNI program. The PPI correlation term described the correlation of the EEG-based regressor with the fMRI time course across all three experimental conditions. The PPI interaction term described the difference in correlations of the EEG-based regressor with the fMRI time course between the Happy Memories condition and the selected control condition (Attend or Count). The fMRI data and motion parameters were bandpass filtered between 0.01 and 0.1 Hz. The GLM design matrix for each task run included four stimulus regressors, ten nuisance covariates, and five polynomial terms for modeling the baseline. The stimulus regressors included: the EEG-based PPI interaction regressor; the EEG-based PPI correlation regressor; the Happy Memories block-stimulus regressor; the Count block-stimulus regressor. The nuisance covariates included: time courses of the six fMRI motion parameters (together with the same time courses shifted by one *TR*); the time course of the WM ROI; the time course of the CSF ROI; the fMRI-visual PPI interaction covariate; the fMRI-visual PPI correlation covariate. The last two nuisance covariates (S1.4) accounted for PPI interaction and correlation effects that could be attributed to temporal variations in the average BOLD activity of the visual cortex V1/V2. Each PPI analysis, conducted in this way, yielded GLM-based $R^2$-statistics and *t*-statistics maps for the EEG-based PPI interaction and correlation terms, which we used to compute voxel-wise PPI interaction and correlation values. The resulting maps were normalized using the Fisher *r*-to-*z* transform, then transformed to the Talairach space, re-sampled to $2\times2\times2$ mm$^3$ isotropic voxel size, and spatially smoothed (5 mm FWHM). See *Supplementary Material* (S1.5) for additional technical details of the PPI analyses.

Independent-samples *t*-tests were conducted to evaluate significance of the EG vs CG group differences in the PPI interaction effects for the Happy vs Attend and the Happy vs Count condition contrasts. The results were corrected for multiple comparisons by controlling the family-wise-error (FWE). The correction was based on Monte Carlo simulations implemented in the AlphaSim AFNI program.

## RESULTS

### Thalamic BOLD Activity

Results of the offline fMRI activation analysis for the AN and MD ROIs are exhibited in Figure 4A for the EG and in Figure 4B for the CG. For statistical testing, each participant's BOLD activity levels were averaged across the four rtfMRI-nf runs (Runs 1-4). The average BOLD activity levels for the Happy Memories conditions (H vs A) for the AN ROI were significant for the EG ($t(14)=5.82$, $p<0.00004$) and showed a significant EG vs CG group difference ($t(27)=3.14$, $p<0.004$). The corresponding average BOLD activity levels for the MD ROI were also significant for the EG ($t(14)=4.86$, $p<0.0003$) and exhibited a significant group difference ($t(27)=2.38$, $p<0.024$). (When results for the two ROIs were tested, the Bonferroni-corrected $p<0.05$ corresponded to uncorrected $p<0.025$). For the Transfer run (TR), the BOLD activity levels (H vs A) for the EG were somewhat lower (Fig. 4A), but did not differ significantly from the average BOLD activity levels across the four rtfMRI-nf runs (AN, TR vs NF: $t(14)=-1.48$, $p<0.162$; MD, TR vs NF: $t(14)=-0.87$, $p<0.402$), indicating transfer of the training effects. Average BOLD activity levels for the LGN positively correlated with those for the MD in either group, as described in *Supplementary Material* (S2.2, Fig. S2).

### Functional Connectivity Changes

Figure 5 shows statistical maps for the EG vs CG group difference in fMRI functional connectivity of the MD during the Happy Memories conditions with rtfMRI-nf. The single-subject functional connectivity analyses included the average fMRI time course of the visual cortex V1/V2 as a nuisance covariate. (The analyses without this covariate provided similar results). Each subject's MD functional connectivity maps were averaged for the last two rtfMRI-nf runs (Runs 3,4, see *Discussion*). Similar fMRI connectivity analyses were conducted for the AN as the seed region. The group difference statistical maps in Fig. 5 were thresholded at $t(27)=\pm2.77$ (uncorr. $p<0.01$) and clusters containing at least 81 voxels were retained to yield FWE corrected $p<0.025$ (to account for testing results for the two seed ROIs). The cluster properties are reported in Table I, separately for the MD and the AN analyses. According to Fig. 5A, the MD functional connectivity with the right medial precuneus (BA 31) was significantly enhanced during the rtfMRI-nf task for the EG compared to the CG. In contrast, the MD functional connectivities with three prefrontal regions and the substantia nigra were significantly reduced for the EG relative to the CG (Figs. 5B-D, Table I).

### Alpha Envelope Correlations

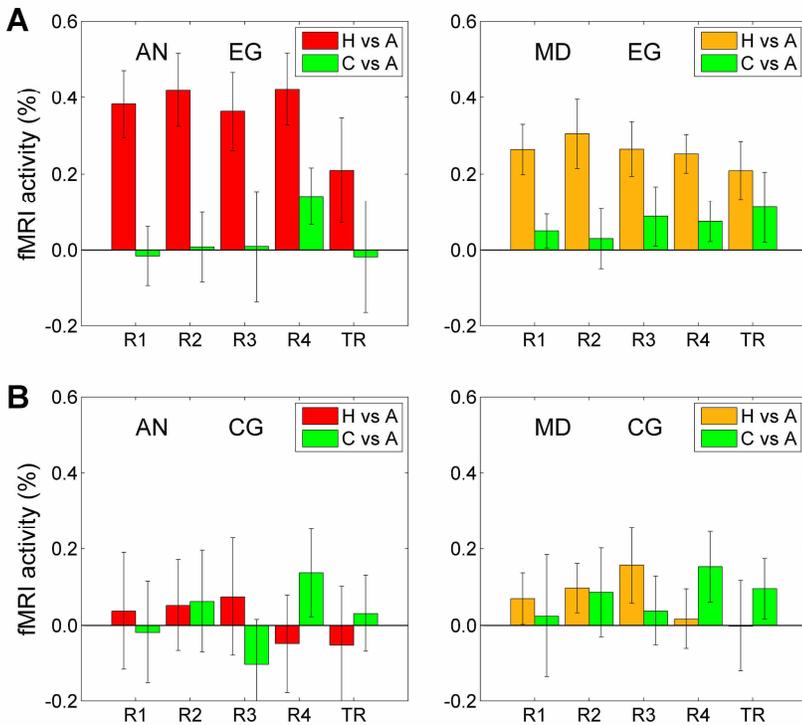

**Figure 4. BOLD activity levels for the anterior and mediodorsal thalamic nuclei during the rtfMRI-nf experiment. A)** Average fMRI percent signal changes for the anterior nucleus (AN) and the mediodorsal nucleus (MD) for the experimental group (EG). The two nuclei were parts of the target ROI for the rtfMRI-nf (Fig. 1A). Each bar represents a mean GLM-based fMRI percent signal change for the corresponding ROI with respect to the Attend baseline for the Happy Memories (H vs A) or Count (C vs A) conditions in a given run, averaged across the group. The error bars are standard errors of the means (sem) for the group average. The experimental runs and condition blocks are depicted schematically in Fig. 2B. **B)** Corresponding average fMRI percent signal changes for the AN and MD for the control group (CG).

Results of the AEC analysis are exhibited in Figure 6. An example of the alpha amplitude envelope is shown in Fig. 6A. The AEC changes between the Happy Memories conditions with rtfMRI-nf and Attend conditions (H vs A) in each task run were computed for each pair of EEG channels. The average AEC change across all channel pairs ($n$=406 for 29 EEG channels) was also computed for each run and each participant. The AEC changes are compared in Figs. 6B-D with fMRI activity levels (marked with *) for the MD ROI corresponding to the part of the total MD BOLD activity that did not exhibit temporal correlation with the average BOLD activity of the visual cortex V1/V2. It was determined from the GLM-based fMRI activation analysis that included the average fMRI time course of the V1/V2 as a nuisance covariate. The average individual AEC changes across all EEG channel pairs (also averaged for Runs 3,4) showed significant positive correlation with the corresponding MD fMRI activity* levels for the EG ($r$=0.56, $p$<0.031, Fig. 6B), but not for the CG (Fig. 6D). Partial correlation between the same measures controlled for the LGN fMRI activity* (Fig. S2) was more positive and significant for the EG ($r(12)$=0.67, $p$<0.008), but not for the CG ($r(11)$=–0.23, $p$<0.446). According to Fig. 6C, the individual AEC changes for the EG showed positive correlations with the MD fMRI activity* levels for many channel pairs (red segments, $r$>0, $p$<0.01, uncorr.). For the CG, no correlations reached the $p$<0.01 statistical threshold. In contrast to the result for the MD in Fig. 6B, correlation between the average AEC changes and fMRI activity* levels for the AN ROI was not significant (AN, EG, NF: $r$=0.23, $p$<0.402).

### Alpha-BOLD Mean Value Correlations

Figure 7 examines correlations between the mean individual changes in the occipital alpha EEG power $z$-scores, $z(\alpha O)$, and the corresponding fMRI activity* levels for the MD ROI. The fMRI activity* has the same meaning as in Fig. 6. The plots on the left show correlation results for the rtfMRI-nf runs (NF, with the data averaged for Runs 3,4). The plots on the right show correlation results for the Transfer run (TR) without nf. According to Fig. 7A, the individual Happy vs Attend (H vs A) occipital alpha power changes for the EG exhibited significant positive correlations with the MD fMRI activity* levels both for the rtfMRI-nf runs (NF: $r$=0.70, $p$<0.004) and for the Transfer run (TR: $r$=0.59, $p$<0.021). No significant correlations were found for the CG (Fig. 7B). Partial correlation between the same measures controlled for the LGN fMRI activity* (Fig. S2) was more significant for the EG (NF: $r(12)$=0.87, $p$<0.0001), but not for the CG (NF: $r(11)$=0.06, $p$<0.848). Compared to the results for the MD in Fig. 7A, correlations between the individual occipital alpha power changes and fMRI activity* levels for the AN ROI were less significant (AN, EG, NF: $r$=0.61, $p$<0.015; TR: $r$=0.29, $p$<0.295).

### Alpha-BOLD Temporal Correlations

Average values of the PPI interaction effects for the AN and MD ROIs are exhibited in Figure 8A. The PPI interaction effects for the Happy vs Attend (H vs A) and Happy vs Count (H vs C) condition contrasts were determined from separate whole-brain PPI analyses using the regressors illustrated in Fig. 3. Voxel-wise PPI interaction values were averaged within the AN and MD ROIs for each run and across the four rtfMRI-nf runs (Runs 1-4) for each participant.

According to Fig. 8A, the average PPI interaction effects for the AN and MD were positive for the EG and negative for the CG. A nonzero PPI interaction indicates that the EEG-based PPI correlation regressor (Fig. 3A) cannot accurately explain the EEG-related variance in the BOLD activity across all conditions, because its correlation with the BOLD time course depends on the condition. Such condition-dependent variance is modeled by the competing EEG-based PPI interaction regressor for a given contrast (Fig. 3C,E). The positive PPI interaction effects for the AN and MD (Fig. 8A) mean that temporal correlations between the occipital alpha EEG power ($z(\alpha O)$, convolved with the HRF) and BOLD activities of these regions were more positive during the Happy Memories conditions with rtfMRI-nf than during the control conditions (Attend or Count) for the EG. These effects are illustrated in Fig. 8B, which compares actual single-subject EEG-based regressor and BOLD activity of the AN across experimental conditions in a single rtfMRI-nf run.

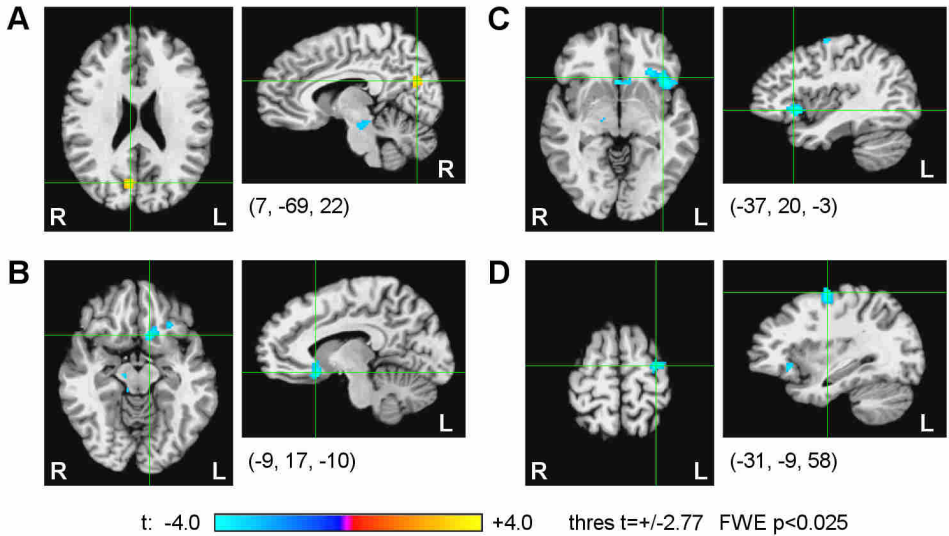

**Figure 5. Functional connectivity of the mediodorsal nucleus during the rtfMRI-nf task compared between the groups.** fMRI functional connectivity analyses were performed for the Happy Memories conditions in each run using the mediodorsal nucleus (MD) ROI as the seed. The results were averaged for the last two rtfMRI-nf runs (Runs 3,4) for each participant. Statistical maps for the experimental vs control group difference (EG vs CG) in the average MD functional connectivity are shown for four regions (Table I): **A)** the right medial precuneus (BA 31); **B)** the left anterior cingulate (BA 32); **C)** the left inferior frontal gyrus (BA 47); **D)** the left precentral gyrus (BA 6). The maps are projected onto the TT_N27 template in the Talairach space. Following the radiological notation, the left hemisphere (L) is shown to the reader's right. The crosshairs mark locations of the statistical peaks for the group difference (Table I).

Importantly, all four EG vs CG group differences in the PPI interaction effects shown in Fig. 8A were significant. (When results for the two ROIs and two contrasts were tested, the Bonferroni-corrected $p<0.05$ corresponded to uncorrected $p<0.0125$). Comparison of the PPI interaction effects for the EG (Fig. 8A) to zero level (instead of the CG) revealed one PPI effect that was significant (EG, AN, H vs C: $t(14)=5.42$, $p<0.0001$), and one PPI effect that trended toward significance after correction (EG, MD, H vs A: $t(14)=2.81$, $p<0.014$). The regression of the fMRI-visual PPI covariates (S1.4) generally improved significance of the PPI results for the AN and MD, as described in *Supplementary Material* (S2.3, Fig. S3).

For the Transfer run for the EG, the mean PPI interaction values (not shown) did not differ significantly from the corresponding PPI interaction values averaged for the four rtfMRI-nf runs (Fig. 8A), suggesting transfer of the training effects. Moreover, the individual PPI interaction values (H vs A) for the MD ROI significantly correlated with the corresponding changes in fMRI functional connectivity strength between the MD and the inferior precuneus, as described in *Supplementary Material* (S2.4, Fig. S4).

Figure 9 shows whole-brain statistical maps for the EG vs CG group difference in the EEG-based PPI interaction effects for the Happy vs Attend condition contrast. The voxel-wise PPI interaction values were averaged across the four rtfMRI-nf runs (Runs 1-4) for each participant. The statistical maps in Fig. 9 were thresholded at $t(27)=\pm2.77$ (uncorr. $p<0.01$) and clusters containing at least 81 voxels were retained (FWE corr. $p<0.025$ to account for testing PPI results for the two contrasts). The cluster properties are described in Table II. The results in Fig. 9 and Table II reveal the largest cluster in the thalamus area (volume ~13 cm$^3$ in the Talairach space), which includes the MD, the AN, the medial pulvinar, and the dorsal parts of the caudate body.

Whole-brain statistical maps for the EG vs CG group difference in the EEG-based PPI interaction effects for the Happy vs Count condition contrast are exhibited in Figure 10. The data were thresholded and clusterized as described above for Fig. 9, and the cluster properties are specified in Table III. The results in Fig. 10 and Table III show the largest cluster (volume ~30 cm$^3$) covering the thalamus area and extending to the dorsal striatum. This cluster includes the AN, the anterior and posterior parts of the MD, the medial pulvinar, the ventrolateral nucleus (VL), the entire caudate body, and the entire putamen.

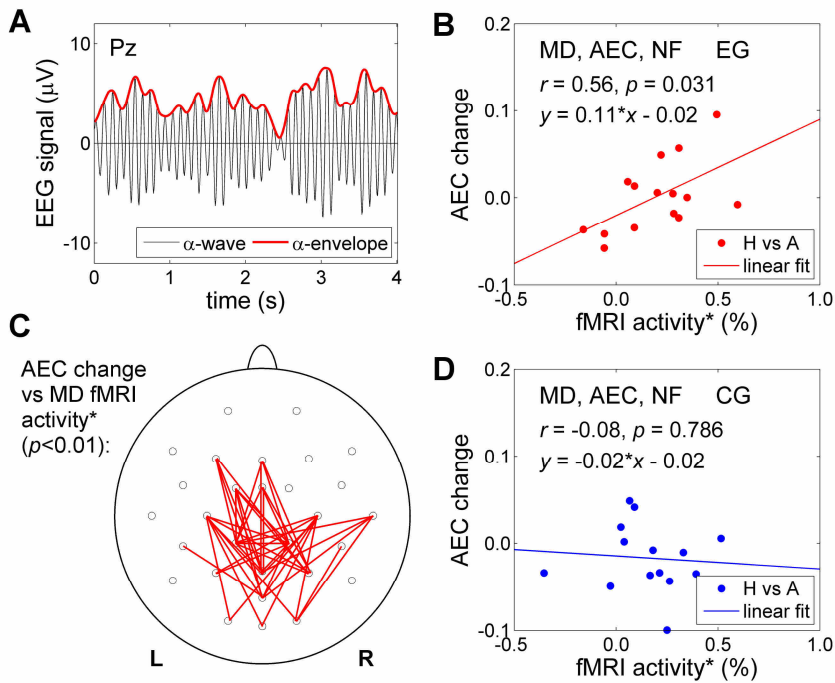

**Figure 6. Amplitude modulation of alpha EEG rhythm during the rtfMRI-nf training and its correlation with BOLD activity of the mediodorsal nucleus. A)** Envelope of alpha EEG activity obtained via the Hilbert transform. Temporal correlation between alpha envelopes for a pair of EEG channels is referred to as the alpha envelope correlation (AEC). **B)** Significant positive correlation between the individual AEC changes for the Happy vs Attend conditions (H vs A), averaged across all EEG channel pairs ($n=406$), and the corresponding fMRI activity* levels (H vs A) for the mediodorsal nucleus (MD) ROI. The results are for the experimental group (EG), and each data point corresponds to one participant. The fMRI activity (marked with *) is the part of the total MD BOLD activity that did not exhibit temporal correlation with the average BOLD activity of the visual cortex V1/V2 (see text). Each participants's data are averaged for the last two rtfMRI-nf runs (Runs 3,4). **C)** Correlations between the individual AEC changes (H vs A) and the corresponding MD fMRI activity* levels (H vs A) for EEG channel pairs for the EG. Each red segment denotes a pair of EEG channels for which the correlation is positive ($r>0$, $p<0.01$, uncorr.). Negative correlations ($r<0$) did not reach the $p<0.01$ statistical threshold. Each participants's data are averaged for Runs 3,4. **D)** Lack of correlation between the individual AEC changes, averaged across all channel pairs, and the corresponding MD fMRI activity* levels for the control group (CG).

We repeated the analyses, reported in Figs. 7-10, using the parietal alpha EEG power (for channels P3, P4, Pz), and they yielded similar results.

## DISCUSSION

In this study, we investigated the feasibility of volitional regulation of BOLD activity of the MD and AN thalamic nuclei using the rtfMRI-nf. We employed simultaneous EEG to evaluate the effects of such non-invasive thalamic modulation on posterior alpha EEG rhythm.

Posterior alpha rhythm is generated by multiple and functionally diverse neuronal sources. MEG equivalent current dipole source localization studies have shown that the strongest dipole-modeled sources of posterior alpha rhythm are distributed around the parieto-occipital sulcus and around the calcarine sulcus [Hari et al., 1997; Manshanden et al., 2002]. The parieto-occipital alpha sources can be located on either side of the sulcus – in the precuneus (BA 7, 31) and in the adjacent occipital cortex (V3, BA 19) [Hari et al., 1997]. Interestingly, positive correlations between global alpha EEG power and glucose metabolism (PET) were found in the same parieto-occipital regions [Schreckenberger et al., 2004]. The parieto-occipital sources of alpha rhythm have been implicated in cognitive and memory functions, including working memory [Tuladhar et al., 2007] and episodic memory [Seibert et al., 2011]. The calcarine alpha sources are located in the visual areas V1/V2 [Portin et al., 1998, 1999]. They are involved in the more basic visual functions, such as pattern detection [Portin et al., 1998] and visual attention [Yamagishi et al., 2003]. Alpha EEG signals measured by the occipital and parietal scalp EEG channels (Fig. 1D) are superpositions of signals from the parieto-occipital and calcarine sources of alpha activity.

In the present study, we focused on correlations between posterior alpha EEG power and BOLD activity of the MD/AN during the autobiographical memory recall. Such correlations are more likely to be associated with the parieto-occipital alpha sources involved in the memory function. Modulation of posterior alpha power by visual attention, associated more closely with the calcarine alpha sources in the visual areas V1/V2, is a confounding factor in our work. Therefore, we took several steps to reduce between- and within-subject variations in visual attention and to account for such variations in the data analysis. *First*, the CG participants were provided with the sham feedback (Fig. 1B), that required visual attention levels similar to those for the rtfMRI-nf (EG). All major conclusions in the study are drawn from the EG vs CG group comparisons (Figs. 4-10). *Second*, the variable-height red rtfMRI-nf bar was displayed during all three experimental conditions (Fig. 2A) to reduce variations in visual attention across the conditions. *Third*, the time course of the occipital alpha EEG power (αO) was orthogonalized with respect to the time course of the eye-blinking activity also reflecting occasional eye closing. *Fourth*, the average fMRI time course of the visual cortex V1/V2 was used to define nuisance covariates for the fMRI activation and functional connectivity analyses, as well as for the PPI analyses of EEG-fMRI correlations. Thus, any effects that could be attributed to temporal variations in the average V1/V2 activity were explicitly regressed out in the GLMs.

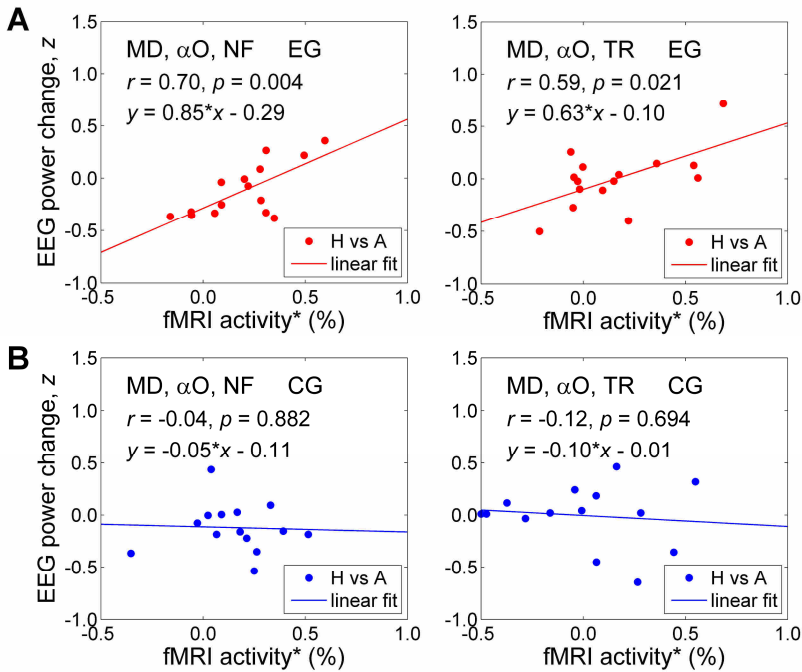

**Figure 7. Correlations between variations in the mean occipital alpha EEG power and BOLD activity of the mediodorsal nucleus during the rtfMRI-nf experiment. A)** Significant positive correlations between the mean individual Happy vs Attend (H vs A) changes in z-scores of the normalized occipital alpha EEG power, $z(\alpha O)$, and the corresponding fMRI activity* levels for the MD ROI for the experimental group (EG). As in Fig. 6, the fMRI activity* is the part of the total MD BOLD activity that did not show temporal correlation with the average BOLD activity of the visual cortex V1/V2. Each data point corresponds to one participant. The plot on the left (NF) shows correlation for the participants' individual data averaged for the last two rtfMRI-nf runs (Runs 3,4). The plot on the right (TR) shows correlation for the Transfer run without nf. **B)** Lack of correlations between corresponding measures for the control group (CG).

During the rtfMRI-nf training, the EG participants were able to successfully upregulate BOLD activity of the target brain region consisting of the AN and MD thalamic nuclei (Fig. 4). The rtfMRI-nf training effects were specific to the EG, and generalized beyond the actual rtfMRI-nf training (Fig. 4A). The MD functional connectivity during the rtfMRI-nf task was enhanced (for the EG compared to the CG) for the right medial precuneus (BA 31), and reduced for several prefrontal regions (Fig. 5, Table I). The functional connectivity between the MD and the precuneus in our study reflects involvement of both regions in the autobiographical memory recall. Importantly, this functional connectivity was enhanced by the rtfMRI-nf for the EG, as evidenced by the significant EG vs CG group difference (Fig. 5A).

Performance of the rtfMRI-nf task was accompanied by amplitude modulation of alpha EEG activity (Fig. 6). The stronger the MD BOLD activity was modulated for the EG, the stronger correlated modulation of alpha amplitude was observed throughout the EEG array (Fig. 6B). We estimate from the linear fit in Fig. 6B that a 1% MD BOLD activity (with the average V1/V2 activity kept constant) would correspond to an increase in the average alpha envelope correlation (AEC) across the array by 0.09. The positive correlation effect for the EG was observed for many channel pairs (Fig. 6C), particularly those involving channels Pz and P4, located over the medial and right lateral precuneus. The correlation pattern in Fig. 6C is consistent with the fMRI functional connectivity results in Fig. 5A and the PPI interaction results in Fig. 9.

During the rtfMRI-nf task, changes in the mean occipital alpha EEG power ($\alpha O$) significantly correlated with the mean MD BOLD activity levels for the EG (Fig. 7A, left). The correlation was also significant for the Transfer run (Fig. 7A, right), during which no visual nf information was provided. Based on the linear fit in Fig. 7A (left), we estimate that a 2.3% MD BOLD activity (with the average V1/V2 activity kept constant) might lead to a significant increase ($z=1.65$, one-tailed $p<0.05$) in the mean occipital alpha power. Importantly, the observed increase in alpha power with increasing MD BOLD activity cannot be attributed to a reduction in visual attention, because the mean LGN BOLD activity levels during the rtfMRI-nf task positively correlated with the mean MD BOLD activity levels (S2.2, Fig. S2).

Overall, our analysis suggests that the more accurately activities of the main visual areas (LGN, V1, V2) are controlled and accounted for, the stronger positive correla-correlations are observed between alpha EEG activity measures (AEC or power changes) and BOLD activity of the MD. For example, the correlation results for the EG, reported in Fig. 6B and Fig. 7A, become more significant if the mean LGN BOLD activity levels are partialled out. Interestingly, significance of the results in Figs. 5A, 6B, 7A increased across the nf runs, likely because the EG participants became more proficient at controlling the rtfMRI-nf signal as the training progressed. Therefore, the average results for the last two nf runs (Runs 3,4) are exhibited in Figs. 5-7. The PPI interaction results in Figs. 8-10 did not show an obvious trend across the nf runs, probably because the PPI results are more strongly affected by residual artifacts and noise in both EEG and fMRI data. Therefore, the average PPI interaction effects across all four nf runs (Runs 1-4) are reported in Figs. 8-10.

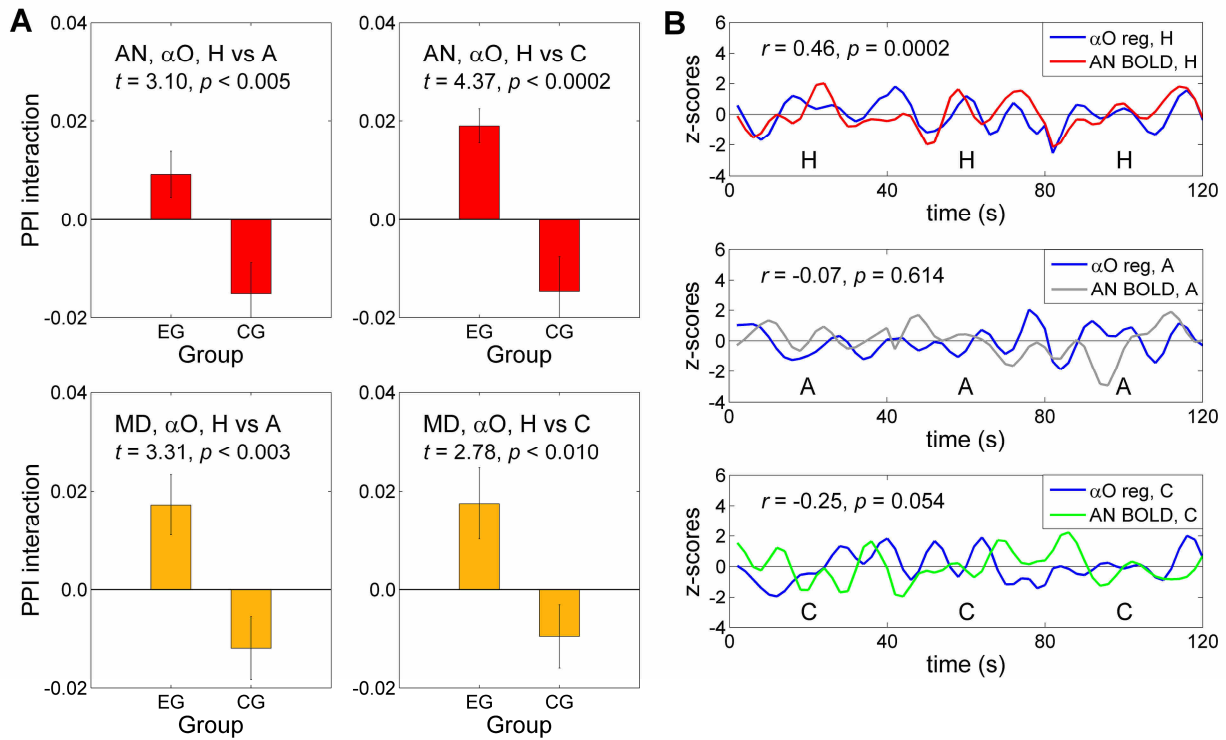

**Figure 8. Changes in temporal correlations between occipital alpha EEG power and BOLD activities of the anterior and mediodorsal nuclei across experimental conditions. A)** Average values of the psychophysiological interaction (PPI) effects for the anterior (AN) and mediodorsal (MD) thalamic nuclei for the experimental group (EG) and the control group (CG). Time courses of *z*-scores of the normalized occipital alpha EEG power, $z(\alpha O)$, were used in the PPI analyses as illustrated in Fig. 3. The EEG-based PPI interaction effect for the Happy vs Attend condition contrast is denoted in the figure as 'H vs A', and the EEG-based PPI interaction effect for the Happy vs Count condition contrast is denoted as 'H vs C'. The voxel-wise PPI interaction values were averaged within the AN and MD ROIs (Fig. 1A) and across the four rtfMRI-nf runs (Runs 1-4) for each participant. Each bar represents a group average. The error bars are standard errors of the means (sem). The *t*-scores and *p*-values in the figure correspond to the EG vs CG group difference ($df=27$). **B)** Illustration of the PPI effects for the EG using single-subject data. The top plot shows positive correlation between the EEG-based regressor and the fMRI time course for the AN ROI during three Happy Memories (H) condition blocks in one nf run (Fig. 2B) concatenated together in the figure. The middle plot demonstrates lack of correlation between these time courses during three concatenated Attend (A) condition blocks, following the Happy Memories blocks in the same run. The bottom plot shows negative correlation between these time courses during three concatenated Count (C) condition blocks in the same run.

The main result of our study is the observation of enhanced temporal correlation between thalamic BOLD activity and alpha EEG rhythm during the rtfMRI-nf training. The positive PPI interaction effects for the AN and MD ROIs for the EG (Fig. 8A) indicate the more positive temporal correlations between the occipital alpha EEG power (convolved with the HRF) and BOLD activities of these nuclei during the rtfMRI-nf task compared to the control tasks. These effects were specific to the EG, and generalized beyond the actual rtfMRI-nf training. Similar PPI effects were observed for the parietal alpha EEG power. Taken together, the results in Figs. 6, 7, and 8 demonstrate both mean value correlations and enhanced temporal correlations between the MD BOLD activity and the modulation of alpha EEG rhythm, suggesting a profound connection between the two processes.

The PPI interaction group-difference results in Fig. 9 and Table II, corresponding to the Happy vs Attend condition contrast, reveal the strongest effect in the right MD at (5, −13, 14). This is consistent with the right lateralization of MD BOLD activity in fMRI autobiographical memory studies [Spreng et al., 2009; Svoboda et al., 2006]. The results in Fig. 9 and Table II also show significant effects in many brain regions known to have connections to the MD, including the dorsolateral prefrontal cortex (DLPFC, specifically BA 9 and BA 6 with $|x|>20$ mm), the ventrolateral prefrontal cortex (BA 45), the anterior cingulate cortex (BA 32), the caudate body, and the cerebellum. The involvement of the DLPFC and the dorsolateral caudate suggests an engagement of the "prefrontal" basal ganglia-thalamocortical circuit [Alexander et al., 1991; Grahn et al., 2008]. This circuit is a closed loop connecting the DLPFC and the parvicellular subdivision of the MD nucleus (MDpc) as follows: DLPFC => caudate => globus pallidus / substantia nigra => MDpc => DLPFC. The main cluster in Fig. 9 also includes the medial pulvinar, which has extensive connections to the prefrontal cortex [Bridge

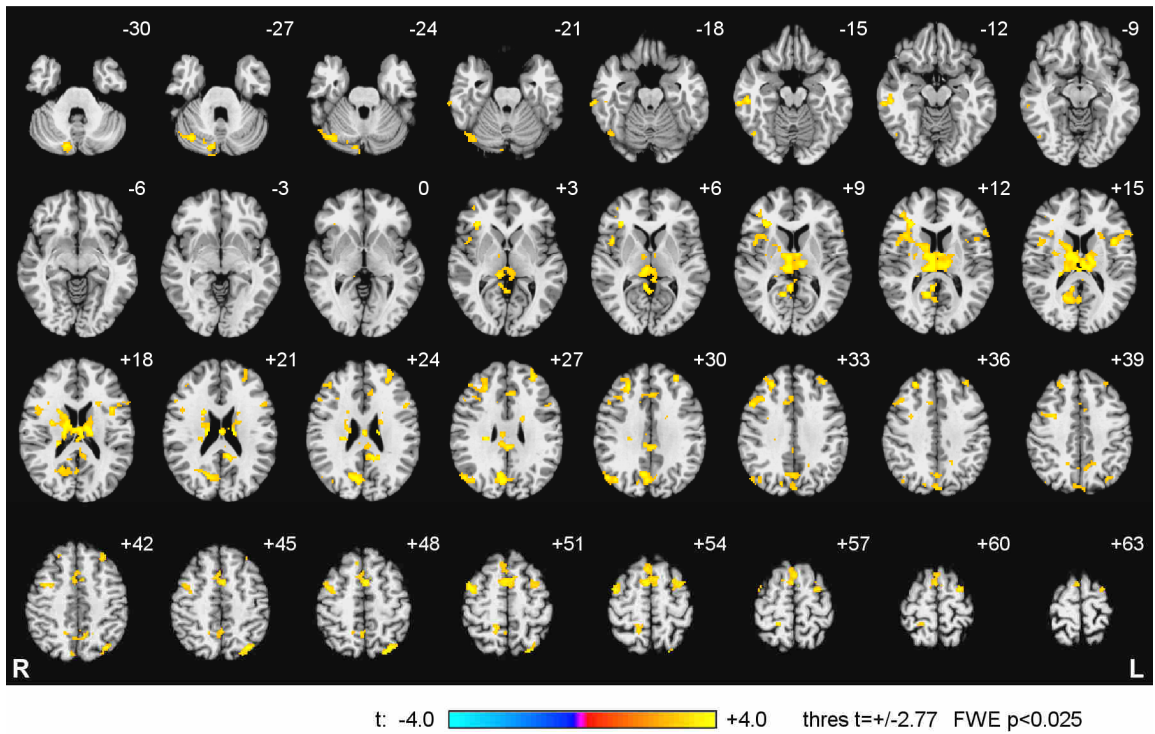

**Figure 9. Enhancement in temporal correlation between occipital alpha EEG power and BOLD activity during the rtfMRI-nf task relative to the Attend task compared between the groups.** Statistical maps for the experimental vs control group difference (EG vs CG) in the EEG-based PPI interaction effect for the Happy vs Attend condition contrast are shown. Time courses of z-scores of the normalized occipital alpha EEG power, $z(\alpha O)$, and the [Happy−Attend] contrast function were used to define the PPI regressors as illustrated in Figs. 3A,B,C. The PPI interaction results for the four rtfMRI-nf runs (Runs 1-4) were averaged for each participant. The maps are projected onto the TT_N27 template in the Talairach space, with 3 mm separation between axial slices. The number adjacent to each slice indicates the $z$ coordinate in mm. The left hemisphere (L) is to the reader's right. Peak $t$-statistics values for the group difference and the corresponding locations are specified in Table II.

et al., 2016]. There are no significant effects in the ventrolateral pulvinar or the LGN.

The second largest cluster in Fig. 9 and Table II has the maximum effect within the right medial precuneus (BA 31) at (6, −70, 25) and extends down to the posterior cingulate cortex (BA 30). As mentioned above, the precuneus/posterior cingulate region is the main posterior node of the DMN. Fig. 9 and Table II also show a cluster in the right angular gyrus (BA 39), closely corresponding to the right lateral DMN node. Thus, the results in Fig. 9 and Table II demonstrate an enhancement in temporal correlation between the occipital alpha EEG power and BOLD activity not only for the thalamus, but also for the posterior (medial and right lateral) DMN nodes. The latter finding is consistent with results of the EEG-only studies that showed increased alpha EEG power within the posterior DMN regions during self-referential processing [Knyazev et al., 2011, 2015]. It is also consistent with the results of the EEG-PET study [Schreckenberger et al., 2004] that showed positive correlations between global alpha EEG power and glucose metabolism under lorazepam challenge in the precuneus (BA 7, 31).

The PPI group-difference results in Fig. 10 and Table III correspond to the Happy vs Count condition contrast. Unlike the Attend condition, the Count condition is the cognitive task that shares some characteristics with the Happy Memories task. It activates the MD (Fig. 4), which is involved in the working memory function, and the parietal regions involved in numerical processing and working memory [Zotev et al., 2011]. Therefore, the PPI results in Fig. 10 and Table III are less sensitive to the general effects of memory performance (particularly related to the MD) and more sensitive to the specific effects of the rtfMRI-nf that distinguish it from the sham feedback. The prominent involvement of the caudate body and the DLPFC confirms the engagement of the "prefrontal" basal ganglia-thalamocortical circuit discussed above. The significant effects in the ventrolateral nucleus (VL), the putamen, and the supplementary motor area (SMA, medial BA 6) indicate an engagement of the "motor" basal ganglia-thalamocortical circuit [Alexander et al., 1991; Grahn et al., 2008]. This circuit is a closed loop connecting the SMA and the VL nucleus pars oralis (VLo) as follows: SMA => putamen => globus pallidus / substantia nigra => VLo => SMA.

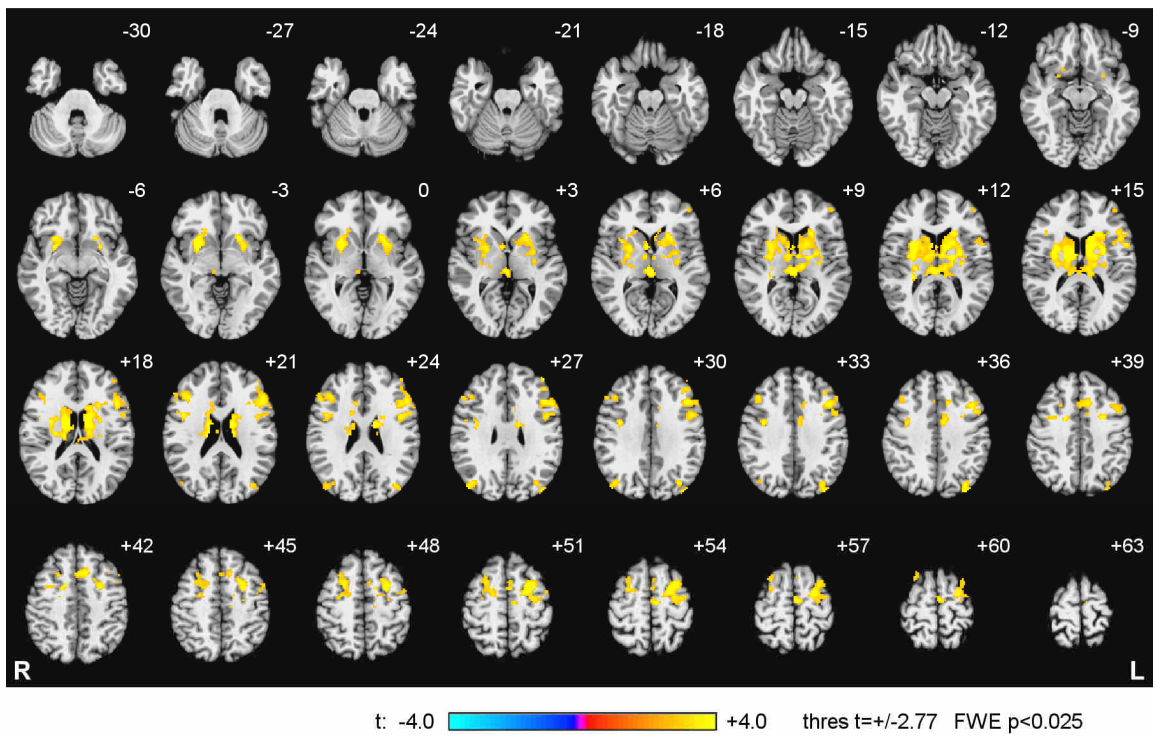

Figure 10. **Enhancement in temporal correlation between occipital alpha EEG power and BOLD activity during the rtfMRI-nf task relative to the Count task compared between the groups.** Statistical maps for the experimental vs control group difference (EG vs CG) in the EEG-based PPI interaction effect for the Happy vs Count condition contrast are shown. Time courses of $z$-scores of the normalized occipital alpha EEG power, $z(\alpha O)$, and the [Happy−Count] contrast function were used to define the PPI regressors as illustrated in Figs. 3A,D,E. The PPI interaction results for the four rtfMRI-nf runs (Runs 1-4) were averaged for each subject. The maps are projected onto the TT_N27 template in the Talairach space, with 3 mm separation between axial slices. The number adjacent to each slice indicates the $z$ coordinate in mm. The left hemisphere (L) is to the reader's right. Peak $t$-statistics values for the group difference and the corresponding locations are specified in Table III.

The results in Fig. 10 and Table III also show a cluster in the right temporo-parietal area (BA 39) corresponding approximately to the right angular gyrus cluster (BA 39) in Fig. 9 and Table II. Another cluster with the maximum effect in the left precuneus (BA 19) has the center of mass at (−34, −76, 32) in the left angular gyrus (BA 39). Therefore, the PPI group-difference results in Fig. 10 and Table III show positive effects in the temporo-parietal regions in the general vicinities of the lateral posterior DMN nodes.

The caudate and the putamen, together constituting the dorsal striatum, serve as the basal ganglia's interface to the cortex. They play important roles in learning, including skill learning and learning from feedback, e.g. [Grahn et al., 2008; Yin & Knowlton, 2006]. Notably, the caudate body and the putamen are associated with the more basic stimulus-response learning [Yin & Knowlton, 2006], while the caudate head is involved in the more advanced action-outcome learning [Grahn et al., 2008]. Self-regulation studies using rtfMRI-nf have shown engagement of these regions [Lawrence et al., 2014; Scharnowski et al., 2015; Sulzer et al., 2013; Veit et al., 2012]. The basal ganglia-thalamocortical circuits, including the "prefrontal" and "motor" circuits mentioned above, enable parallel processing and integration [Alexander et al., 1991], and play crucial roles in the learning cycle [Birbaumer et al., 2013; Yin & Knowlton, 2006]. The results in Fig. 10 and Table III demonstrate that BOLD activities of these circuits' regions correlate with the posterior alpha EEG activity. This finding is not surprising, because these circuits are engaged during the rtfMRI-nf procedure targeting the MD BOLD activity, which positively correlates with alpha EEG power. However, the correlation effects in the caudate and putamen (Fig. 10) may also reflect a more profound connection between oscillations in the basal ganglia and in the cortex, e.g. [Brittain & Brown, 2014]. For example, oscillatory connectivity in the alpha band has been observed between the subthalamic nucleus and temporo-parietal cortical regions [Litvak et al., 2011].

Our results suggest that the functional connectivity between the MD and the inferior precuneus (BA 31) plays an important role during the described rtfMRI-nf procedure. Such connectivity was enhanced during the rtfMRI-nf task for the EG compared to the CG for the parietal regions, with the statistical maximum at (7, −69, 22) within the right medial precuneus (Fig. 5A, Table I). This locus is spatially close to the parieto-occipital sulcus (Fig. 5A). Modulation of this region's BOLD activity (together with that of the MD) may conceivably affect, directly or indirectly, the main dipolar sources of alpha EEG rhythm in the vicinity of the parieto-occipital sulcus, discussed above. Indeed, temporal correlation between the occipital alpha EEG power and BOLD activity was enhanced during the rtfMRI-nf task for the EG relative to the CG for the posterior DMN nodes, with the maximum at (6, −70, 25) within the right medial precuneus (Fig. 9, Table II). This location is remarkably close to the above-mentioned point of the strongest MD connectivity enhancement. Furthermore, the PPI interaction effect for the MD significantly correlated, for the Transfer run for the EG, with the enhancement in the MD-precuneus functional connectivity (S2.4, Fig. S4A). Taken together, these findings suggest that the temporal correlation between the MD BOLD activity and posterior alpha EEG power is modulated by the interaction between the MD and the inferior precuneus, reflected in their functional connectivity.

The rtfMRI-nf targeting the MD and/or AN, provided during a memory task (episodic, working, spatial, etc), can conceivably be used for the memory function training. Such training may be relevant in neuropsychiatric disorders characterized by autobiographical memory disturbances, including depression and PTSD. For example, the rtfMRI-nf can be used to increase BOLD activity of the MD during recall of happy autobiographical memories and/or reduce such activity in response to traumatic memories or rumination. The ability of the rtfMRI-nf to alter functional connectivity between the MD and the posterior DMN (Fig. 5A) may be particularly important in treatment of depression, which is characterized by abnormally elevated resting fMRI connectivity between these regions [Greicius et al., 2007; Hamilton et al., 2015].

Our findings suggest that EEG-nf based on modulation of alpha EEG power would naturally complement the rtfMRI-nf targeting the MD/AN. Studies utilizing EEG-nf for upregulation of alpha (or upper alpha) EEG power have demonstrated improvements in participants' cognitive and memory performance, including working memory and short term memory, e.g. [Escolano et al., 2014; Hanslmayr et al., 2005; Hsueh et al., 2016; Nan et al., 2012; Zoefel et al., 2011]. Changes in resting fMRI functional connectivity following alpha EEG-nf training have also been reported [Kluetsch et al., 2014; Nicholson et al., 2016]. Interestingly, most alpha EEG-nf studies did not employ specific cognitive tasks or predefined mental strategies during the training. Using alpha EEG-nf during a particular cognitive or memory task would conceivably improve training of the related function compared to the non-specific training scenario. For example, upregulation of posterior alpha EEG power using EEG-nf during recall of happy autobiographical memories could specifically enhance episodic memory of emotional events. Such EEG-nf approach can be used as a supplementary technique in combination with the rtfMRI-nf of the MD/AN. The two methods can also be used simultaneously as rtfMRI-EEG-nf [Zotev et al., 2014] to enable simultaneous regulation of thalamic BOLD activity and alpha EEG rhythm.

## CONCLUSION

We demonstrated, for the first time, that healthy participants can learn to successfully upregulate BOLD activities of the mediodorsal and anterior thalamic nuclei using the rtfMRI-nf. We observed that such regulation enhanced temporal correlation between thalamic BOLD activity and alpha EEG power. We also identified the posterior DMN nodes as cortical regions supporting this correlation effect. Our results confirm the fundamental role of the thalamus, particularly the mediodorsal nucleus, in modulation of alpha EEG rhythm. They also demonstrate the potential of the rtfMRI-nf with simultaneous EEG for non-invasive neuromodulation studies of human brain function.

## ACKNOWLEDGMENTS

This work was supported by the Laureate Institute for Brain Research and the William K. Warren Foundation. We would like to thank Dr. Patrick Britz, Dr. Robert Störmer, Dr. Florian Strelzyk, and Dr. Mario Bartolo of Brain Products, GmbH for their help and technical support. The authors have no conflict of interests to declare.

**Table I. Functional connectivities of the mediodorsal and anterior nuclei during the rtfMRI-nf task compared between the groups.** Statistical results for the experimental vs control group differences (EG vs CG) in fMRI functional connectivities of the mediodorsal nucleus (MD) and anterior nucleus (AN) during the Happy Memories conditions with rtfMRI-nf are shown. The functional connectivity analyses were performed separately for the MD and the AN as seed regions. Location of the point with the peak group difference $t$-score ($df$=27) and the number of voxels are specified for each cluster obtained after FWE correction for multiple comparisons in each analysis.

| Region | Laterality | x, y, z (mm) | $t$-score | Size (vox.) |
|---|---|---|---|---|
| **Mediodorsal nucleus** | | | | |
| Inferior frontal gyrus (BA 47) | L | −37, 20, −3 | −4.84 | 217 |
| Anterior cingulate (BA 32) | L | −9, 17, −10 | −4.80 | 131 |
| Precentral gyrus (BA 6) | L | −31, −9, 58 | −4.35 | 100 |
| Substantia nigra | R | 11, −17, −8 | −3.92 | 98 |
| Precuneus (BA 31) | R | 7, −69, 22 | +3.86 | 98 |
| **Anterior nucleus** | | | | |
| Caudate head | L | −15, 19, 4 | −4.60 | 217 |
| Lentiform nucleus | L | −31, −5, −6 | −4.62 | 82 |

BA – Brodmann areas; L – left; R – right; x, y, z – Talairach coordinates;
FWE corrected $p<0.025$ (Size – cluster size, minimum 81 voxels for uncorr. $p<0.01$).

**Table II. Enhancement in temporal correlation between occipital alpha EEG power and BOLD activity during the rtfMRI-nf task relative to the Attend task compared between the groups.** Statistical results for the experimental vs control group difference (EG vs CG) in the EEG-based PPI interaction effect for the Happy vs Attend condition contrast are shown. Location of the point with the peak group difference *t*-score (*df*=27) and the number of voxels are specified for each cluster obtained after FWE correction for multiple comparisons. For the largest cluster in the thalamus/dorsal striatum area, local *t*-score maxima are also specified.

| Region | Laterality | x, y, z (mm) | t-score | Size (vox.) |
|---|---|---|---|---|
| **Thalamus / dorsal striatum** | | | | |
| Mediodorsal nucleus (MD) | R | 5, –13, 14 | 5.81 | 1591 |
| Mediodorsal nucleus (MD) | L | –5, –21, 15 | 4.59 | –//– |
| Caudate body | L | –15, –17, 22 | 5.49 | –//– |
| Caudate body | R | 16, –16, 20 | 4.83 | –//– |
| Pulvinar, medial | R | 5, –23, 10 | 4.87 | –//– |
| **Frontal lobe** | | | | |
| Superior frontal gyrus (BA 6) | R | 5, 5, 54 | 4.47 | 465 |
| Inferior frontal gyrus (BA 45) | R | 33, 27, 5 | 5.09 | 388 |
| Middle frontal gyrus (BA 9) | R | 25, 37, 34 | 4.50 | 348 |
| Middle frontal gyrus (BA 6) | R | 39, –3, 52 | 4.46 | 213 |
| Superior frontal gyrus (BA 9) | L | –29, 43, 28 | 4.14 | 179 |
| Middle frontal gyrus (BA 6) | L | –25, –3, 62 | 4.51 | 116 |
| **Limbic lobe** | | | | |
| Cingulate gyrus (BA 23) | L | –3, –33, 28 | 3.82 | 115 |
| Posterior cingulate (BA 30) | L | –7, –43, 22 | 4.84 | 97 |
| Anterior cingulate (BA 32) | L | –7, 27, 25 | 3.62 | 96 |
| **Parietal lobe** | | | | |
| Precuneus (BA 31) / posterior cingulate | R | 6, –70, 25 | 4.75 | 732 |
| Precuneus (BA 7) | L | –28, –69, 48 | 4.63 | 150 |
| Angular gyrus (BA 39) | R | 47, –67, 30 | 3.99 | 147 |
| Precuneus (BA 7) | R | 3, –53, 48 | 3.36 | 96 |
| **Temporal lobe** | | | | |
| Inferior temporal gyrus (BA 20) | R | 63, –27, –16 | 4.28 | 122 |
| **Other** | | | | |
| Insula (BA 13) | L | –39, 7, 16 | 4.35 | 217 |
| Declive (cerebellum) | R | 37, –67, –22 | 4.01 | 167 |
| Pyramis (cerebellum) | R | 13, –77, –30 | 4.18 | 115 |

BA – Brodmann areas; L – left; R – right; x, y, z – Talairach coordinates;
FWE corrected $p<0.025$ (Size – cluster size, minimum 81 voxels for uncorr. $p<0.01$);
–//– denotes a local maximum within the same cluster.

**Table III. Enhancement in temporal correlation between occipital alpha EEG power and BOLD activity during the rtfMRI-nf task relative to the Count task compared between the groups.** Statistical results for the experimental vs control group difference (EG vs CG) in the EEG-based PPI interaction effect for the Happy vs Count condition contrast are shown. Location of the point with the peak group difference *t*-score (*df*=27) and the number of voxels are specified for each cluster obtained after FWE correction for multiple comparisons. For the largest cluster in the thalamus/dorsal striatum area, local *t*-score maxima are also specified.

| Region | Laterality | x, y, z (mm) | t-score | Size (vox.) |
|---|---|---|---|---|
| **Thalamus / dorsal striatum** | | | | |
| Pulvinar, medial | L | −1, −27, 6 | 7.41 | 3757 |
| Ventrolateral nucleus (VL) | L | −18, −9, 12 | 6.81 | –//– |
| Caudate body | R | 11, −5, 18 | 6.17 | –//– |
| Caudate body | L | −13, −5, 22 | 6.13 | –//– |
| Putamen | R | 27, 3, 14 | 5.28 | –//– |
| Putamen | L | −27, −3, 14 | 5.05 | –//– |
| Anterior nucleus (AN) | R | 9, −7, 13 | 5.04 | –//– |
| Anterior nucleus (AN) | L | −8, −5, 13 | 4.35 | –//– |
| **Frontal lobe** | | | | |
| Middle frontal gyrus (BA 9) | L | −41, 19, 32 | 5.23 | 980 |
| Middle frontal gyrus (BA 6) | L | −29, 3, 52 | 5.63 | 736 |
| Middle frontal gyrus (BA 9) | R | 35, 25, 22 | 4.14 | 250 |
| Medial frontal gyrus (BA 6) | L | −5, −13, 58 | 4.76 | 91 |
| **Limbic lobe** | | | | |
| Cingulate gyrus (BA 24) | R | 17, 1, 42 | 4.99 | 580 |
| Cingulate gyrus (BA 32) | L | −9, 17, 42 | 4.34 | 193 |
| **Parietal lobe** | | | | |
| Precuneus (BA 19) | L | −31, −79, 38 | 4.75 | 194 |
| **Temporal lobe** | | | | |
| Middle temporal gyrus (BA 39) | R | 41, −73, 28 | 3.92 | 102 |

BA – Brodmann areas; L – left; R – right; *x, y, z* – Talairach coordinates;
FWE corrected $p<0.025$ (Size – cluster size, minimum 81 voxels for uncorr. $p<0.01$);
–//– denotes a local maximum within the same cluster.

## SUPPLEMENTARY MATERIAL

### S1.1. fMRI Data Analysis

Offline analysis of the fMRI data was performed in AFNI [Cox, 1996; Cox & Hyde, 1997]. Pre-processing of single-subject fMRI data included correction of cardiorespiratory artifacts using the AFNI implementation of the RETROICOR method [Glover et al., 2000]. Further fMRI pre-processing involved slice timing correction and volume registration of all EPI volumes acquired in the experiment using the 3dvolreg AFNI program with two-pass registration.

The fMRI activation analysis was performed using the standard general linear model (GLM) approach. It was conducted for each of the five task fMRI runs (Fig. 2B) using the 3dDeconvolve AFNI program. The GLM model included two block-design stimulus condition terms, Happy Memories and Count (Fig. 2B), represented by the standard block-stimulus regressors in AFNI. A general linear test term was included to compute the Happy vs Count contrast. Nuisance covariates included the six fMRI motion parameters and five polynomial terms for modeling the baseline. GLM $\beta$ coefficients were computed for each voxel, and average percent signal changes for Happy vs Attend, Count vs Attend, and Happy vs Count contrasts were obtained by dividing the corresponding $\beta$ values ($\times 100\%$) by the $\beta$ value for the constant baseline term. The resulting fMRI percent signal change maps for each run were transformed to the Talairach space by means of the @auto_tlrc AFNI program using each subject's high-resolution anatomical brain image as the template.

Average individual BOLD activity levels for the AN and MD thalamic nuclei were computed in the offline analysis for the AN and MD ROIs, exhibited in Fig. 1A. The ROIs were defined anatomically as specified in the AFNI implementation of the Talairach-Tournoux brain atlas (TT_N27). The voxel-wise fMRI percent signal change data from the GLM analysis, transformed to the Talairach space, were averaged within the AN and MD ROIs and used as GLM-based measures of these regions' BOLD activities. In addition, we determined average individual BOLD activity levels for the LGN ROI, also defined anatomically based on the TT atlas in AFNI.

The fMRI functional connectivity analyses for the MD and AN as the seed regions were performed within the GLM framework. The fMRI data were bandpass filtered between 0.01 Hz and 0.1 Hz. The six fMRI motion parameters were similarly filtered. The MD ROI (Fig. 1A) was transformed to each subject's individual high-resolution anatomical image space, and then to the individual EPI image space. The MD ROI in the EPI space included ~200 voxels. In addition, bilateral 10-mm-diameter ROIs were defined within white matter (WM) and ventricle cerebrospinal fluid (CSF) and similarly transformed. The resulting ROIs in the individual EPI space were used as masks to obtain average time courses for the MD, WM, and CSF regions. The GLM-based functional connectivity analysis was conducted for each task run using the 3dDeconvolve AFNI program. The -censor option was used to restrict the analysis to the Happy Memories condition blocks in each run. The GLM model included the time course of the MD ROI as the stimulus (seed) regressor. Nuisance covariates included five polynomial terms, time courses of the six fMRI motion parameters (together with the same time courses shifted by one $TR$), time courses of the WM and CSF ROIs to suppress physiological noise [Jo et al., 2010], and step functions to account for the breaks in the data between the Happy Memories condition blocks. Each GLM analysis provided $R^2$-statistics and $t$-statistics maps for the stimulus regressor term, which we used to compute the correlation coefficient for each voxel. The correlation coefficient maps were Fisher $r$-to-$z$ normalized, transformed to the Talairach space, and re-sampled to $2\times2\times2$ mm$^3$ isotropic voxel size. The resulting individual MD functional connectivity maps were spatially smoothed (5 mm FWHM) and submitted to group analyses. The fMRI functional connectivity analysis for the AN as the seed region was conducted in a similar way.

In addition to the described standard fMRI activation and functional connectivity analyses, we performed similar analyses with the average fMRI time course of the visual cortex V1/V2 included as a nuisance covariate. The V1/V2 ROI was defined anatomically as the combination of Brodmann areas 17 (V1) and 18 (V2), specified in the AFNI implementation (TT_N27) of the Talairach-Tournoux brain atlas. The resulting ROI was transformed to each subject's individual EPI space and used as a mask to obtain the average fMRI time course for V1/V2. The GLM analyses with this additional covariate yielded fMRI results corresponding to partial BOLD activities that did not exhibit temporal correlations with the average BOLD activity of the V1/V2.

### S1.2. EEG Data Processing

Removal of MR and cardioballistic (CB) artifacts was based on the average artifact subtraction method implemented in BrainVision Analyzer 2.1 (Brain Products, GmbH). The MR artifact template was defined using MRI slice markers recorded with the EEG data. After the MR artifact removal, the EEG data were bandpass filtered between 0.5 and 80 Hz (48 dB/octave) and downsampled to 250 S/s sampling rate (4 ms interval). The fMRI slice selection frequency (17 Hz) and its harmonics were removed by band rejection filtering. The CB artifact template was determined from the cardiac waveform recorded by the ECG channel, and the CB artifact to be subtracted was defined, for each channel, by a moving average over 21

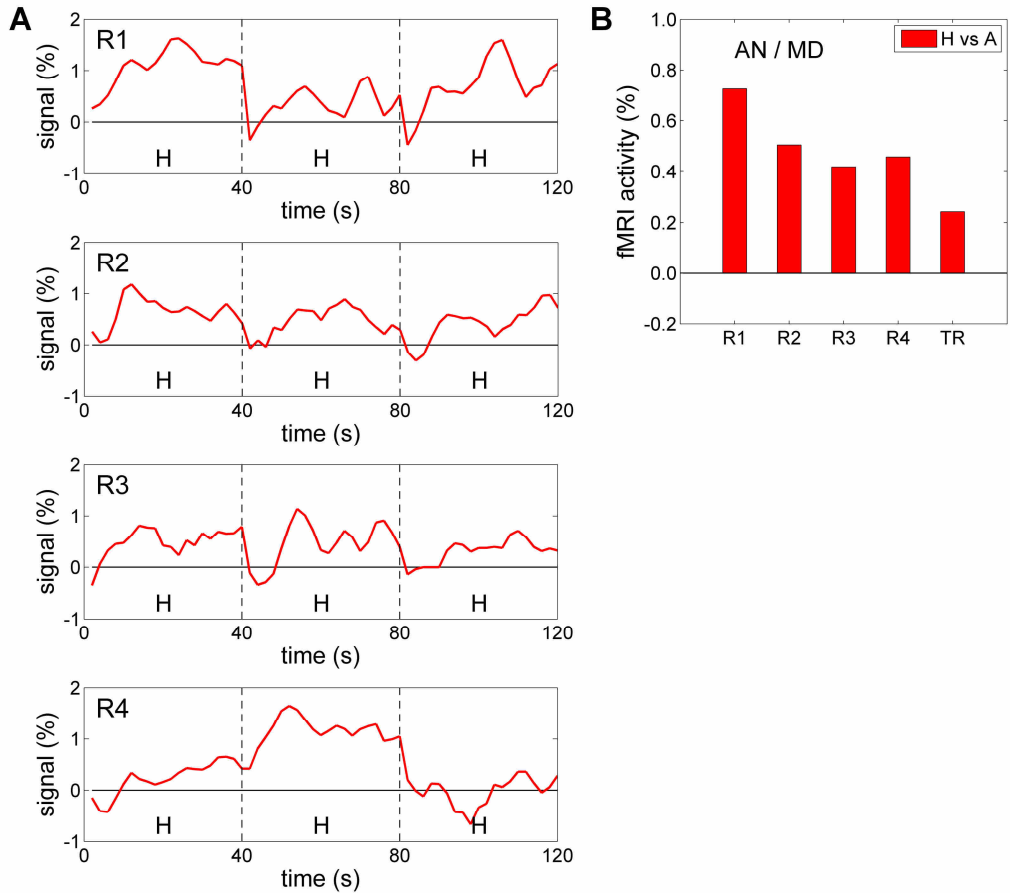

**Figure S1. Real-time fMRI neurofeedback signal for a single subject. A)** Time courses of the actual rtfMRI-nf signal presented to one EG participant during Happy Memories (H) condition blocks in the four rtfMRI-nf training runs (R1-R4). Each condition block was 40 s long and included acquisition of 20 fMRI volumes. The rtfMRI-nf signal was updated every *TR*=2 s. The three Happy Memories condition blocks in each run are concatenated together in the figure. **B)** Mean values of these rtfMRI-nf signals across the Happy Memories conditions in each run. For the Transfer run (TR), the signal was obtained in the same way as for the preceding rtfMRI-nf runs, but was not shown to the participant.

cardiac periods. Intervals with strong motion artifacts were not included in the CB correction.

Following the MR and CB artifact removal, the EEG data from the five task runs (Fig. 2B) were concatenated to form a single dataset. The data were carefully examined, and intervals exhibiting significant motion or instrumental artifacts ("bad intervals") were excluded from the analysis. Channel Cz was selected as a new reference, and FCz was restored as a regular channel.

An independent component analysis (ICA) was performed over the entire dataset with exclusion of the bad intervals. This approach ensured that independent components (ICs) corresponding to various artifacts were identified and removed in a consistent manner across all five runs. Channels TP9 and TP10 were excluded from the ICA and further analysis, because their signals are very sensitive to head, jaw, and ear movements, producing large artifacts. The Infomax ICA algorithm (Bell & Sejnowski, 1995), implemented in BrainVision Analyzer 2.1, was applied to the data from 29 EEG channels and yielded 29 ICs. Time courses, spectra, topographies, and kurtosis values of all the ICs were carefully analyzed (see e.g. McMenamin et al., 2010 and supplement therein) to identify various artifacts, as well as EEG signals of neuronal origin, with particular attention to the alpha and theta EEG bands. After all the ICs had been classified, an inverse ICA transform was applied to remove the identified artifacts from the EEG data. Following the ICA-based artifact removal, the EEG data were low-pass filtered at 40 Hz (48 dB/octave) and downsampled to 125 S/s (8 ms interval). Because many artifacts had been already removed using the ICA, the data were examined again, and new bad intervals were defined to exclude remaining artifacts.

### S1.3. EEG Eye-blinking Covariates

To take into account possible modulating effects of eye movements [de Munck et al., 2007], including eye blinking and occasional eye closing, on the alpha EEG power, we defined an EEG eye-blinking covariate for each run as follows. Eye-blinking activity was approximated by back-projection of the ICs, describing eye-blinking artifacts, onto the frontopolar EEG channels Fp1 and Fp2.

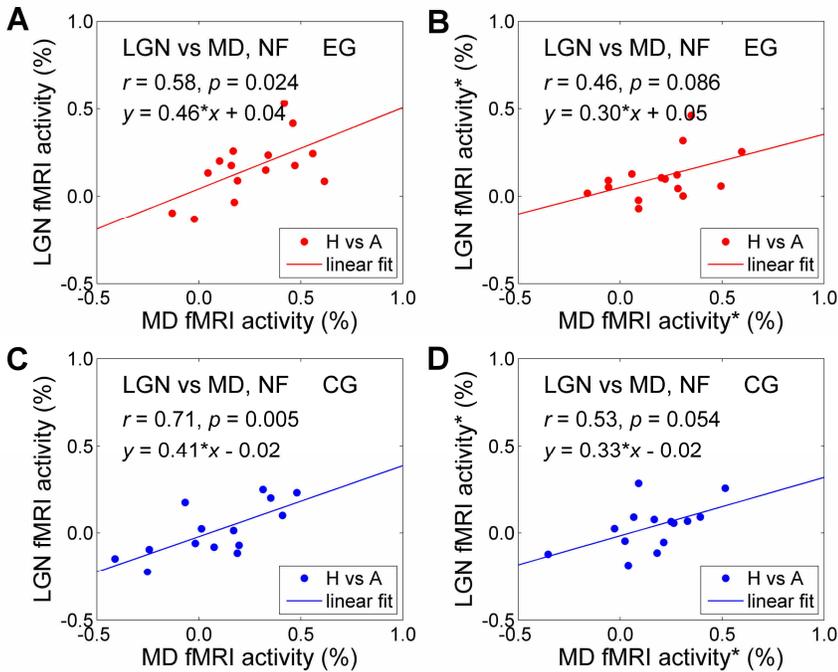

**Figure S2. Correlations between BOLD activity levels for the mediodorsal nucleus and the lateral geniculate nucleus during the rtfMRI-nf training.** The individual GLM-based fMRI activity levels for the Happy Memories conditions with respect to the Attend baseline (H vs A) for the MD and LGN ROIs were averaged for Runs 3,4. Each data point corresponds to one participant. **A)** Correlation between the average MD and LGN fMRI activity levels for the experimental group (EG). **B)** Same as in A), but the fMRI activity levels (marked with *) were obtained from GLM analyses with the average fMRI time course of the visual cortex V1/V2 included as a nuisance covariate. **C)** Correlation between the average MD and LGN fMRI activity levels for the control group (CG). **D)** Same as in C), but the fMRI activity* levels were obtained from GLM analyses with the average fMRI time course of the visual cortex V1/V2 included as a nuisance covariate.

The continuous wavelet transform was performed to compute signal power for these two channels. Normalized power values at each time point were defined as $\ln(P(Fp1)) + \ln(P(Fp2))$, where $P$ is the eye-blinking signal power in the low frequency range [0.25...5.0] Hz. The power values were averaged for 200-ms long time bins, linearly detrended, and converted to $z$-scores across each run. The resulting covariate was regressed out from the corresponding time course of the $z$-scores of the normalized occipital alpha EEG power, $z(\alpha O)$.

### S1.4. fMRI-visual PPI Covariates

In addition to the EEG-based PPI regressors, we defined fMRI-based PPI regressors using the average fMRI time course of the visual cortex V1/V2. These regressors were defined in the way that is standard for an fMRI-only PPI analysis. The average V1/V2 fMRI time course, bandpass filtered between 0.01 and 0.1 Hz, was used as a PPI correlation regressor, and we refer to it as fMRI-visual PPI correlation covariate. A PPI interaction regressor was defined as follows. The average V1/V2 fMRI time course, bandpass filtered between 0.01 and 0.1 Hz, was detrended with respect to similarly filtered time courses of the six fMRI motion parameters, the WM and CSF ROIs using the 3dDetrend AFNI program. It was then deconvolved using the 3dTfitter AFNI program to estimate the time course of the underlying neuronal activity. This 'neuronal' time course was multiplied by a selected contrast function, [Happy−Attend] or [Happy−Count], and convolved with the same HRF ('Cox special') using the waver AFNI program. We refer to the resulting waveform as fMRI-visual PPI interaction covariate (for a given contrast). We used these fMRI-visual PPI covariates as nuisance covariates in the PPI analyses of the EEG-fMRI data.

### S1.5. Additional Details of PPI Analyses

Three separate PPI analyses were conducted for each fMRI task run. One analysis included the EEG-based PPI interaction regressor (Fig. 3C) corresponding to the [Happy−Attend] contrast function shown in Fig. 3B. For the second analysis, an alternative [Happy−Attend] contrast function (not shown) was defined as +1 for the Happy Memories condition blocks, −1 for the *preceding* Attend condition blocks, and 0 for all other points. A new EEG-based PPI interaction regressor was computed using this alternative [Happy−Attend] contrast function, and a similar PPI analysis was performed. The PPI interaction maps from the two analyses were averaged to yield a single resulting map of the PPI interaction effect for the Happy vs Attend condition contrast. The third PPI analysis included the EEG-based PPI interaction regressor (Fig. 3E) corresponding to the [Happy−Count] contrast function shown in Fig. 3D. In each of these analyses, the fMRI-visual PPI interaction covariate (S1.4) was defined using the same contrast function as the EEG-based PPI interaction regressor.

### S2.1. Example of rtfMRI Neurofeedback Signal

Figure S1 shows time courses of the actual rtfMRI-nf signal presented to one EG participant during Happy Memories condition blocks in the four rtfMRI-nf training runs (Runs 1-4). The rtfMRI-nf signal values were based on rtfMRI activity of the AN/MD target ROI (Fig. 1A). They were computed as a moving average of the current and two preceding rtfMRI percent-signal-change values for the target ROI.

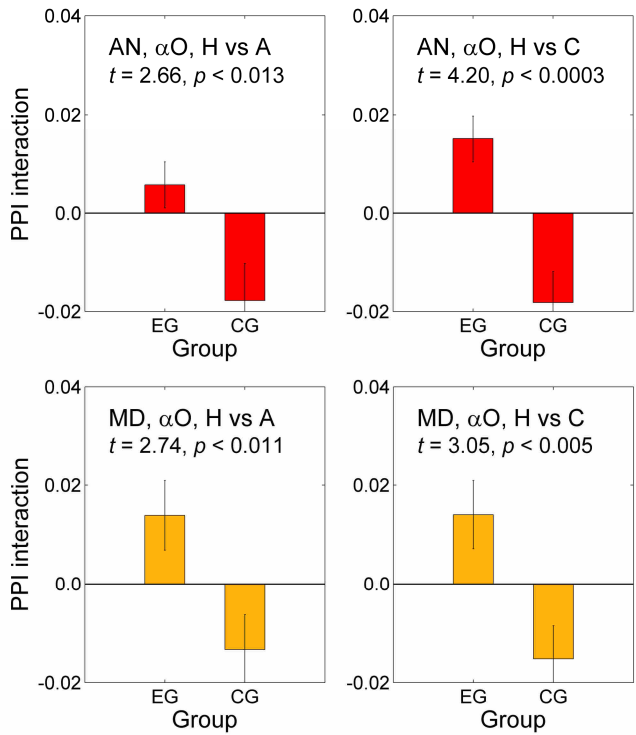

**Figure S3. Psychophysiological interaction (PPI) results for the anterior and mediodorsal nuclei, obtained without regression of the fMRI-visual PPI covariates.** The notations are the same as in Figure 8A of the main text. The results were obtained from similar PPI analyses, but without inclusion of the fMRI-visual PPI correlation and interaction covariates (described in S1.4) as GLM nuisance covariates.

### S2.2. Comparison of MD and LGN BOLD Activities

Figure S2 illustrates correlations between average individual BOLD activity levels for the MD and LGN ROIs during the rtfMRI-nf task. The LGN vs MD fMRI activity correlations are positive and significant for both the EG (Fig. S2A) and the CG (Fig. S2C). When the average fMRI time course of the V1/V2 is explicitly regressed out in the GLM analysis, the LGN vs MD fMRI activity* correlations are also positive and trending toward significance for both the EG (Fig. S2B) and the CG (Fig. S2D). These results suggest that, on the average, the MD and LGN nuclei activated together during the rtfMRI-nf task, though the LGN activity was lower than that of the MD. Importantly, there is no indication that activation of the MD was accompanied by de-activation of the LGN, based on their average BOLD activity levels. Moreover, the LGN vs MD activity correlations were quite similar for the EG and the CG.

### S2.3. Effects of fMRI-visual PPI Covariates

To evaluate effects of the regression of the fMRI-visual PPI covariates, we repeated the PPI analyses described in the main text (EEG-fMRI Temporal Correlation Analysis) without including the fMRI-visual PPI correlation and interaction covariates (S1.4) as nuisance covariates in the GLM models. The results are exhibited in Figure S3. They should be compared to the corresponding PPI results (obtained with the regression) in Fig. 8A. The comparison shows that the regression of the fMRI-visual PPI covariates improves significance of the EG vs CG group differences in the PPI interaction effects (Fig. 8A vs Fig. S3) in three cases out of four. It also improves significance of the average PPI interaction effects for the EG (see Alpha-BOLD Temporal Correlations). The strongest PPI interaction effect for the AN in Fig. 8A (EG, AN, H vs C: $t(14)=5.42$, $p<0.0001$) is more significant than the corresponding effect in Fig. S3 (EG, AN, H vs C: $t(14)=3.19$, $p<0.007$). The strongest PPI interaction effect for the MD in Fig. 8A (EG, MD, H vs A: $t(14)=2.81$, $p<0.014$) is more significant than the corresponding effect in Fig. S3 (EG, MD, H vs A: $t(14)=1.98$, $p<0.067$). Importantly, the EG vs CG group differences in the PPI interaction effects without the regression of the fMRI-visual PPI covariates are also significant (Fig. S3). This means that these PPI effects are real, and not induced by the regression of the fMRI-visual PPI covariates.

### S2.4. Functional Connectivity of MD and Precuneus

The role of fMRI functional connectivity between the MD and the inferior precuneus in examined in Figure S4. The precuneus was defined anatomically as specified in the AFNI implementation of the TT brain atlas, and its lower half with $z \leq 36$ mm was used as a bilateral inferior precuneus ROI (PCun). Voxel-wise fMRI functional connectivity values, obtained from the GLM functional connectivity analysis with the MD seed ROI (S1.1), were averaged for the PCun ROI to yield the average fMRI connectivity ('MD-PCun') between the MD and the inferior precuneus. Voxel-wise values of the EEG-based PPI interaction effect for the Happy vs Attend condition contrast were averaged separately for the MD ROI and for the PCun ROI.

According to Fig. S4A, the PPI interaction effect for the Happy vs Attend (H vs A) condition contrast for the MD ROI showed significant positive correlation ($r=0.64$, $p<0.010$) with the enhancement in the fMRI functional connectivity between the MD and the inferior precuneus ('MD-PCun') during the Transfer run for the EG. The corresponding PPI interaction effect for the PCun ROI exhibited a similar positive correlation ($r=0.65$, $p<0.008$), according to Fig. S4B. The correlations were also significant if the MD-PCun fMRI connectivity strengths during the Happy Memories conditions were included in the analyses instead of the connectivity changes (MD: $r=0.65$, $p<0.009$; PCun: $r=0.66$, $p<0.007$). Thus, the stronger functional connectivity between the MD and the inferior precuneus was associated with the stronger PPI interaction effects for both regions. For the rtfMRI-nf runs for the EG, correlations between the average individual MD-PCun fMRI functional connectivity changes and the corresponding PPI interaction values were also positive, but less significant. This was likely due to stronger modulatory effects of visual attention during the rtfMRI-nf training runs. For the CG, no significant correlations were found between the fMRI connectivity changes and the PPI interaction values (Figs. S4C,D).

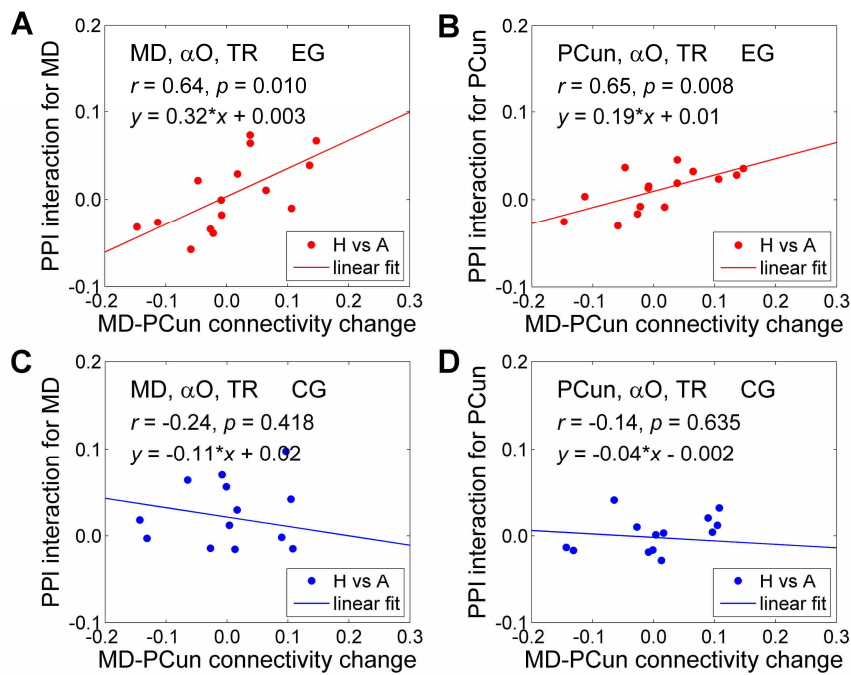

**Figure S4. Functional connectivity between the mediodorsal nucleus and the inferior precuneus and its relation to the PPI interaction effect.** Average individual values of fMRI functional connectivity ('MD-PCun') between the MD ROI (Fig. 1A) and the inferior half of the precuneus (PCun, $z \leq 36$ mm) were computed, and their changes between the Happy Memories conditions and the Attend baseline (H vs A) in the Transfer run (TR) were considered. The fMRI connectivity changes were compared to the corresponding individual EEG-based PPI interaction values for the Happy vs Attend condition contrast, averaged separately for the MD ROI and for the PCun ROI. **A)** Correlation between the fMRI connectivity changes and the PPI interaction values for the MD ROI for the experimental group (EG). **B)** Correlation between the fMRI connectivity changes and the PPI interaction values for the PCun ROI for the EG. **C)** Lack of correlation between the fMRI connectivity changes and the PPI interaction values for the MD ROI for the control group (CG). **D)** Lack of correlation between the fMRI connectivity changes and the PPI interaction values for the PCun ROI for the CG.